\documentclass[namedreferences,hyperref,optionalrh,solaromanenum]{spr-sola}

\usepackage{graphicx}

\graphicspath{{./figures/}}

\usepackage{amssymb}

\usepackage{amsmath}

\usepackage{color}

\usepackage{breakurl}

\usepackage{cancel}

\usepackage{array}

\usepackage{siunitx}

\chardef\us=`\_

\usepackage{tikz}

\DeclareUnicodeCharacter{2605}{$\star$}

\DeclareUnicodeCharacter{2009}{\,}

\newcommand*{\circled}[1]{\tikz[baseline=(char.base)]{

\node[shape=circle, draw, inner sep=1pt] (char) {#1};}}

\begin{document}

\begin{frontmatter}

\title{Augmentation of the Gauribidanur radioheliograph for observations of circularly polarized emission from the solar corona at low radio frequencies}

\author[addressref=aff1,corref,email={shaik.sayuf@iiap.res.in}]{\inits{Sayuf}\fnm{Sayuf}~\snm{Shaik}\orcid{0000-0003-4598-6830}}

\author[addressref={aff1},email={kathir@iiap.res.in}]{\inits{Kathiravan}\fnm{Kathiravan}~\snm{C}\orcid{0000-0002-6126-8962}}

\author[addressref={aff2},email={gireesh@iiap.res.in}]{\inits{Gireesh}\fnm{Gireesh}~\snm{G.V.S}\orcid{0000-0003-0741-0144}}

\author[addressref={aff2},email={indrajit@iiap.res.in}]{\inits{Indrajit}\fnm{Indrajit V.}~\snm{Barve}\orcid{0000-0002-1030-9082}}

\author[addressref={aff2},email={rajesh.m@iiap.res.in}]{\inits{Rajesh}
\fnm{Rajesh}~\snm{M}}

\author[addressref={aff2},email={battini.shamukha@iiap.res.in}]{\inits{Shanmukha}\fnm{Shanmukha}~\snm{B}}

\author[addressref={aff1},email={ramesh@iiap.res.in}]{\inits{Ramesh}\fnm{Ramesh}~\snm{R}\orcid{0000-0003-2651-0204}}



\address[id=aff1]{Indian Institute of Astrophysics, Koramangala 2nd Block, Bangalore 560034, Karnataka, India}

\address[id=aff2]{Radio Astronomy Field Station, Indian Institute of Astrophysics, Gauribidanur 561210, Karnataka, India}

\runningauthor{Sayuf et al.}

\runningtitle{Solar coronal polarization observations at low radio frequencies}

\begin{abstract}
The magnetic field strengths ($B$) in the solar photosphere are routinely measured using the Zeeman effect. However, such measurements are not available in the solar chromosphere and corona. The Gauribidanur RAdioheliograPH (GRAPH), near Bangalore in India, has been successfully carrying out total intensity (Stokes I) observations of the solar corona at low frequencies (${\lesssim}$\,100\,MHz) for several years now. Radio emission from the Sun in the above frequency range originates typically in the heliocentric distance $r\,{\gtrsim}\,1.2\,R_\odot$. 
We recently augmented GRAPH with the aim to observe circularly polarized radio emission (Stokes V) from solar corona, in particular, from the `quiet' corona.
Combining the low frequency Stokes I \& V radio images, and modeling techniques, it is possible to calculate the magnetic field strengths in the solar corona on regular basis. This work details the procedures undertaken to augment the GRAPH for simultaneous observations of both Stokes I \& V radio emission from the solar corona, and the calibration procedure. Preliminary results indicate that the augmented GRAPH can effectively detect Stokes V emission associated with thermal emission from the `quiet' Sun at frequencies ${\lesssim}$\,100\,MHz, in addition to Stokes V emission associated with the non-thermal radio bursts from the `active' Sun.
\end{abstract}



\end{frontmatter}




\section{Introduction}\label{sec:intro}

The Michelson Doppler Imager (MDI: \citealp{scherrer_1995}) onboard the SOlar and Heliospheric Observatory (SOHO), and Helioseismic Magentic Imager (HMI; \citealp{schou_2012}) onboard the Solar Dynamics Observatory (SDO) provide continuous observations of the solar photospheric magnetic field. Similar to MDI and HMI, several ground based radio observatories, some in the past and others at present, have reported several interesting results related to the magnetic field strengths in the solar chromosphere and corona \citep{Kruger_1993ASPC,Lin2000,Lin2004,Gelfreikh_2004ASSL,Ryabov2004,White2004,Aliss_2021}. Results from studies conducted at low radio frequencies ($\lesssim$\,100\,MHz), thus far, indicate that 
the magnetic field strengths are in the range of $\approx$\,0.1-15\,G \citep{dulk_1978corona,vrsnak_2002AA,Ramesh_2005AA,Sastry_2009ApJ,Tun_2013ApJ,Hari_2014ApJ,Kishore_2016ApJ,Kumari_2017SoPh,Kumari_2019ApJ,ramesh_2021thermalCME,Ramesh_2022ApJ,Ramesh_2023ApJ}. Majority of the data underlying these findings were obtained from radio spectroscopic observations of solar radio bursts. The different types of non-thermal radio burst phenomena observed in these spectral data have higher flux densities and measuring their degree of circular polarization (DCP, ratio of circularly polarized intensity to total intensity) is also relatively easier. Compared to the radio bursts, the DCP associated with thermal radio emission from the solar corona are much lesser. Even though, detecting this DCP from thermal emission will enable us to estimate the coronal magnetic field strengths on regular basis than the occasional solar radio bursts. Note that, although the thermal bremsstrahlung emission is intrinsically unpolarized, it can show a small amount of circular polarization in the presence of a magnetic field in the coronal plasma due to the differential absorption of ordinary and extra-ordinary rays in the birefringent magnetized coronal plasma. Since the emission is weak, the measurements present a significant challenge. Observations at frequencies $<$\,100\,MHz offer certain advantages in this connection because of radio wave propagation characteristics like refraction due to density gradients in the solar corona and relatively lesser optical depth when compared to higher frequencies (see, e.g. \citealp{smerd1950,Alissandrakis1994a,Golap_&_Sastry_(1994),Ramesh_2005a,Sastry_2009ApJ}). So, we augmented the existing GRAPH array \citep{Ramesh_1998,Ramesh_2014} which is a dedicated array for solar radio observations and operating primarily at frequencies $<$100\,MHz, to observe circularly polarized radio emission (Stokes V) from the Sun in addition to its present capability of only total intensity (Stokes I) observations. In this work, we present the details of the existing GRAPH, the augmented GRAPH, data calibration, preliminary observations and results, and future plans.

\section{The GRAPH (before augmentation)}\label{sec:graph}
\begin{figure}
\centering
\includegraphics[width=1.0\textwidth]{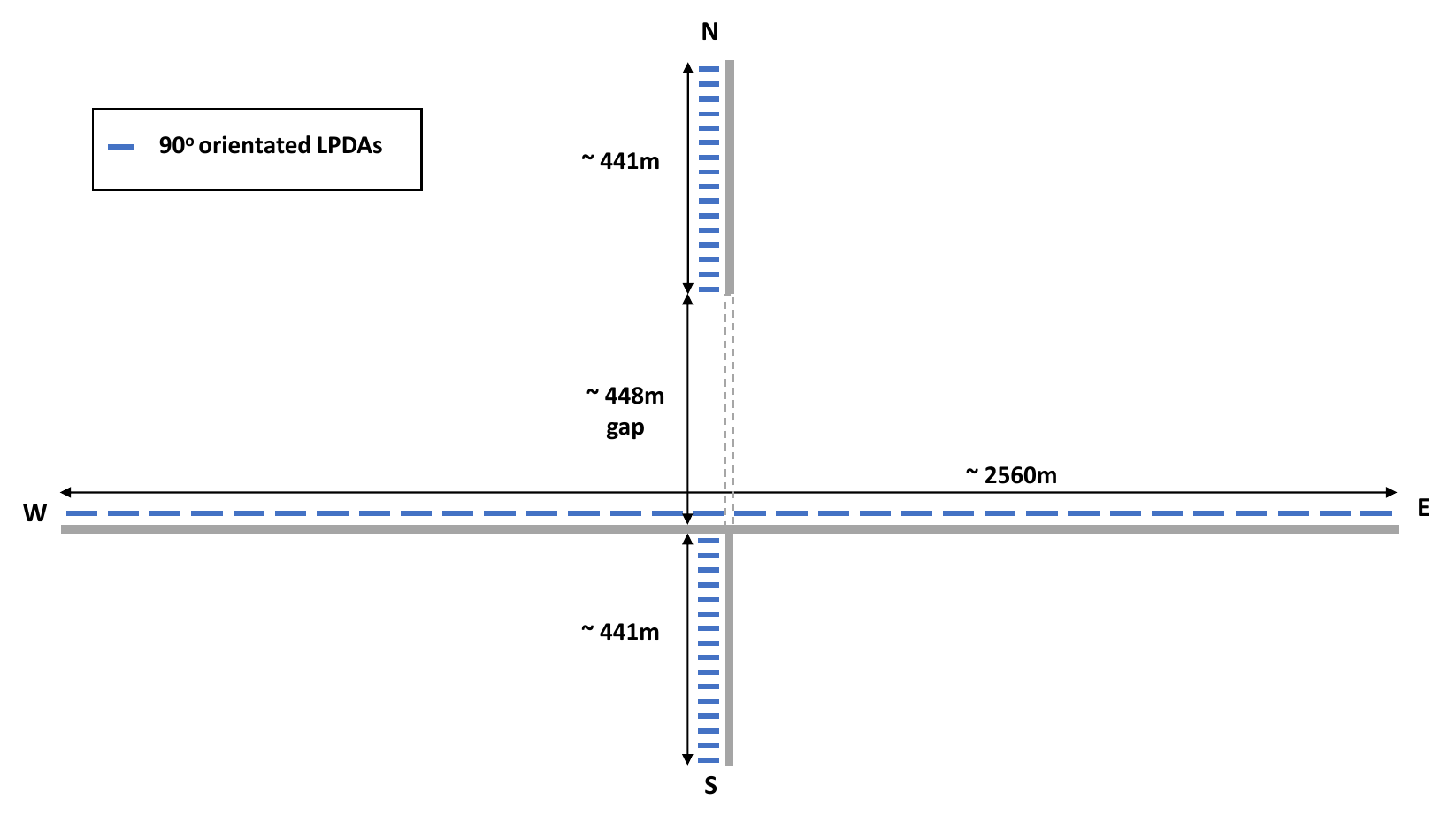}
\caption{Layout of the GRAPH array (not to scale), before augmentation. Each blue line in the EW arm represents group of 8 LPDAs. In the NS arm, each blue line  represents groups of 4 LPDAs. All the LPDAs are \ang{90} orientation w.r.t the celestial north.}
\label{fig:layout_pre}
\end{figure}

GRAPH is a two-dimensional radio interferometer array for dedicated observations of the solar corona which can operate in the 40\,-\,150\,MHz frequency range. It has been operating at the Gauribidanur Radio Observatory (GRO; Longitude: \ang{77.44} E \& Latitude: \ang{13.60} N), situated about 100 km North West of Bangalore (Karnataka, India). Prior to augmentation, the array consisted of 128 antennas each in the East and West directions, and 64 antennas each in the North and South directions. The total number of LPDAs in the array was 384. The basic receiving element used is a Log Periodic Dipole Array (LPDA) antenna. The spacing between the adjacent LPDAs is 10\,m in the East and West arms, and 7\,m in the North and South arms. The 1st LPDA in the South arm is in line with the LPDAs in the East-West arm. The location of the above mentioned LPDA is considered as the center of array for reference purposes. The arrangement gives a total length of 2560\,m for the array in the East-West (EW) direction. The length of the South arm is 441\,m and 1st LPDA in the North arm starts from a distance of 448\,m from the center of the array. In principle, all the 64 LPDAs in the North arm should have been present in the South arm, extending it, after the 64th antenna there. But we couldn't do so because of space constraints and hence the above arrangement. Considering symmetry \citep{Christiansen1987}, the effective baseline length in the North-South (NS) direction of the array is 2${\times}$889\,m (Figure \ref{fig:layout_pre}). 

\begin{figure}
\centering
\includegraphics[width=0.8\textwidth]{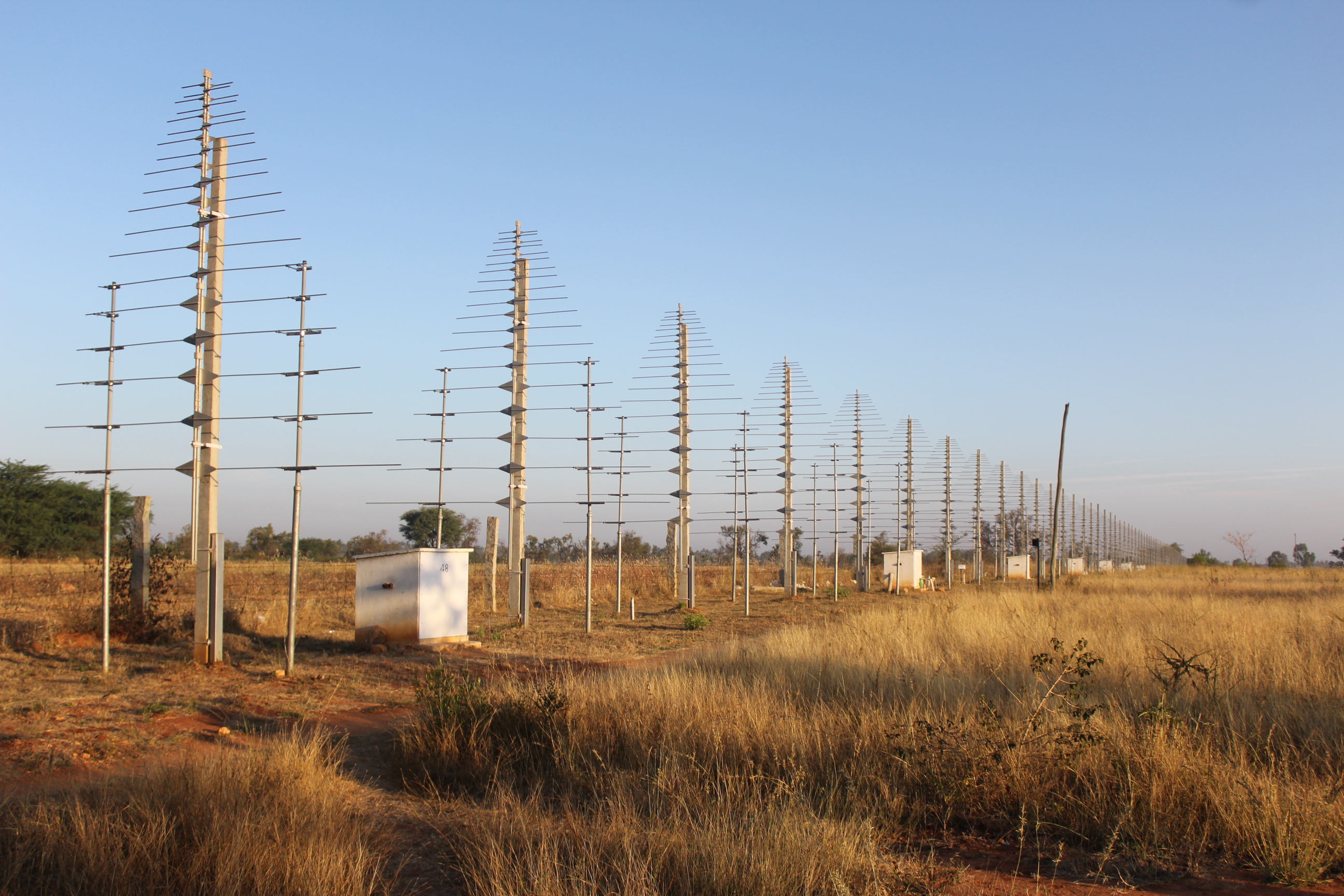}
\caption{Photograph showing a section of the GRAPH array south arm, before augmentation.}
\label{fig:southarm_pre}
\end{figure}

A combination of eight adjacent LPDAs in the EW arm form one group. In the NS arm, four LPDAs form a group. A section of the South arm is shown in Figure \ref{fig:southarm_pre}. The total number of antenna groups in the array is 64. The inter-group spacing in the EW and NS arms are 80\,m, and 28\,m, respectively.
The white-painted kiosks in Figure \ref{fig:southarm_pre}, located at the center of each group, house the front-end receiver modules. Each LPDA can observe in the 40\,-\,150 MHz frequency band, and has an effective collecting area of $\approx$0.5\,$\lambda^2$. It's Voltage Standing Wave Ratio (VSWR) is $\lesssim$\,2.0 (or impedance $\approx 50\,\Omega$) in the above frequency range, and the half-power beam width (HPBW) is ${\approx}$\,\ang{80} in the E-plane (along the dipole plane), and ${\approx}$\,\ang{110} in the H-plane (perpendicular to the dipole plane). As shown in Figure \ref{fig:southarm_pre}, the antennas are mounted vertically and point towards the local zenith. To prevent sagging, the antenna arms are supported by non-conducting structures. Since the LPDA is a large structure and the array has 384 of them, physically steering them to track any celestial source in hour angle and any desired declination is practically difficult. To observe celestial sources whose declinations are different from the local zenith, the antennas are electronically steered towards the direction of the source using a Declination Control Unit (DCU) which is an analog beamformer. 
The DCUs used in the GRAPH can steer the beam across declination in the steps of \ang{8}. The LPDA is a linearly polarized antenna, i.e. it can receive radio waves whose electric field oscillations are parallel to its dipoles. The dipoles in all the LPDAs in GRAPH are oriented along the EW direction (\ang{90} orientation or x-direction as per IAU\citealp{IAU_1973} convention). The reference orientation is the celestial North (\ang{0}). The overall characteristics and performance of the GRAPH, before augmentation, are listed in Table \ref{tab:GRAPH}. 
\begin{table}[h]
\caption{GRAPH – specifications and performance}
\label{tab:GRAPH}
\begin{tabular}{ll}
\hline
\textbf{Parameter} & \textbf{Value} \\
\hline
Basic receiving element & Log periodic dipole array (LPDA) antenna \\
Frequency of operation & 40\,-\,150\,MHz \\
LPDA - HPBW & approx. $\rm \ang{80} \times \ang{120} $ ($\rm \theta_E \times \phi_H$) \\
Polarization & Linear (along the dipole arm/E-plane) \\
Declination coverage & \ang{-31} S to \ang{+59} N \\
Observing period & 02:00-10:00 UT \\
Number of antenna groups & 64 (32 EW and 32 NS) \\
Inter-group spacing & 80 m (EW arm) \& 28 m (NS arm) \\
LPDAs per group & 8 (EW) \& 4 (NS) \\
Total number of LPDAs & 384 \\
Effective collecting area & 192~$\rm \lambda^2$ \\
Field of view & $\ang{1.5} \times \ang{4}$ at 150 MHz \\
Angular resolution & $2.7^{\prime} \times 4.0^{\prime}$ (RA $\times$ Dec.) at 150 MHz \\
Resolution Bandwidth & 1 MHz \\
Sampling rate & 4 MHz \\
Correlation receiver & 4096 channels (1 bit - 2 level) \\
Temporal resolution & 256 ms \\
Sensitivity (theoretical) & $\approx$\,2\,-\,3\,Jhansky at 150 MHz for 10 s integration time \\
Dynamic range of images & $\approx$\,22\,dB \\
\hline
\end{tabular}
\end{table}

\begin{figure}
\centering
\includegraphics[width=0.98\textwidth]{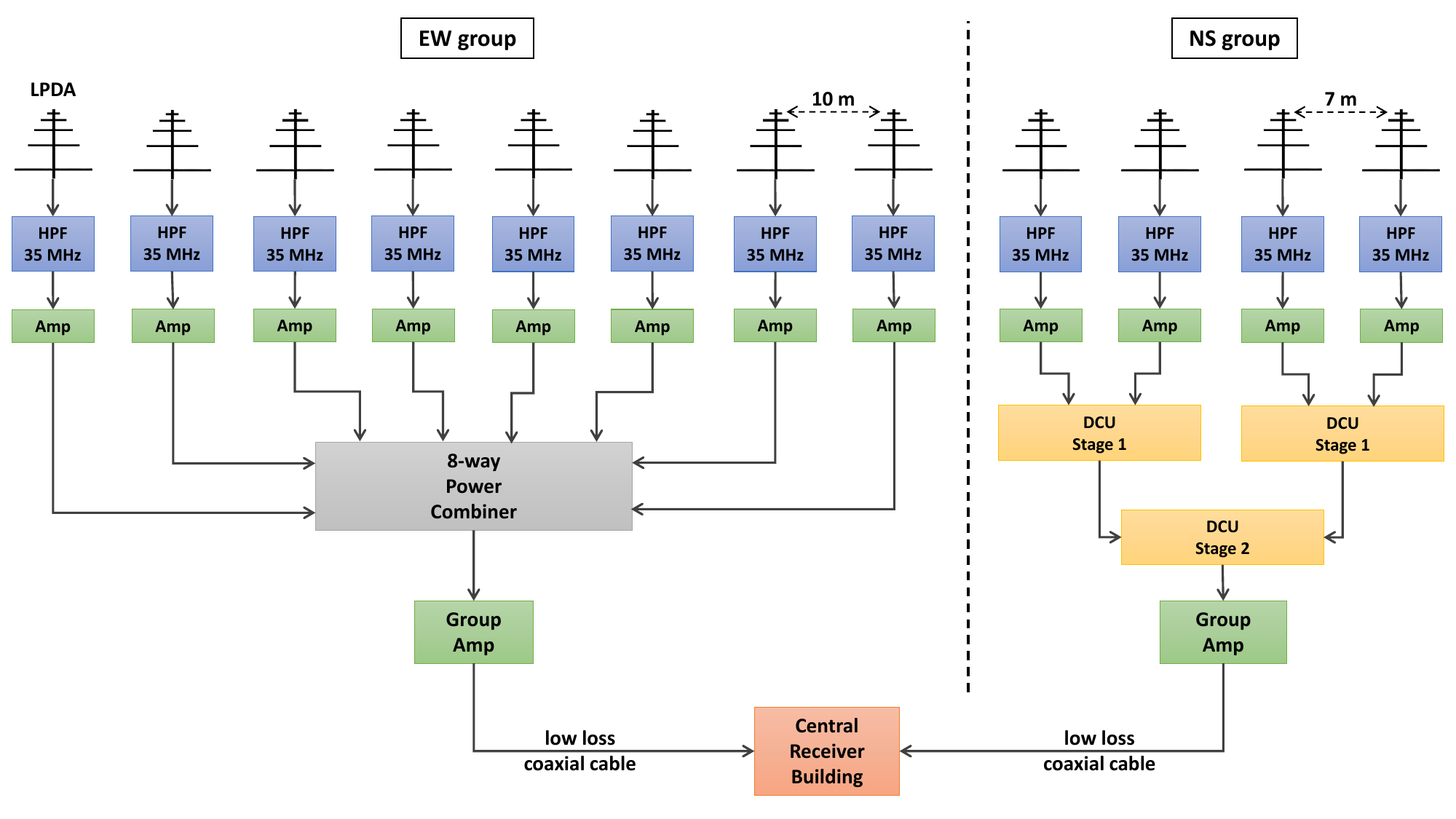}
\caption{Block diagram of an analog front-end receiver system of the GRAPH EW (left) and NS (right) groups at the antenna array site, before augmentation.}
\label{fig:Afrontend_graph_pre}
\end{figure}

\begin{figure}
\centering
\includegraphics[width=0.98\textwidth]{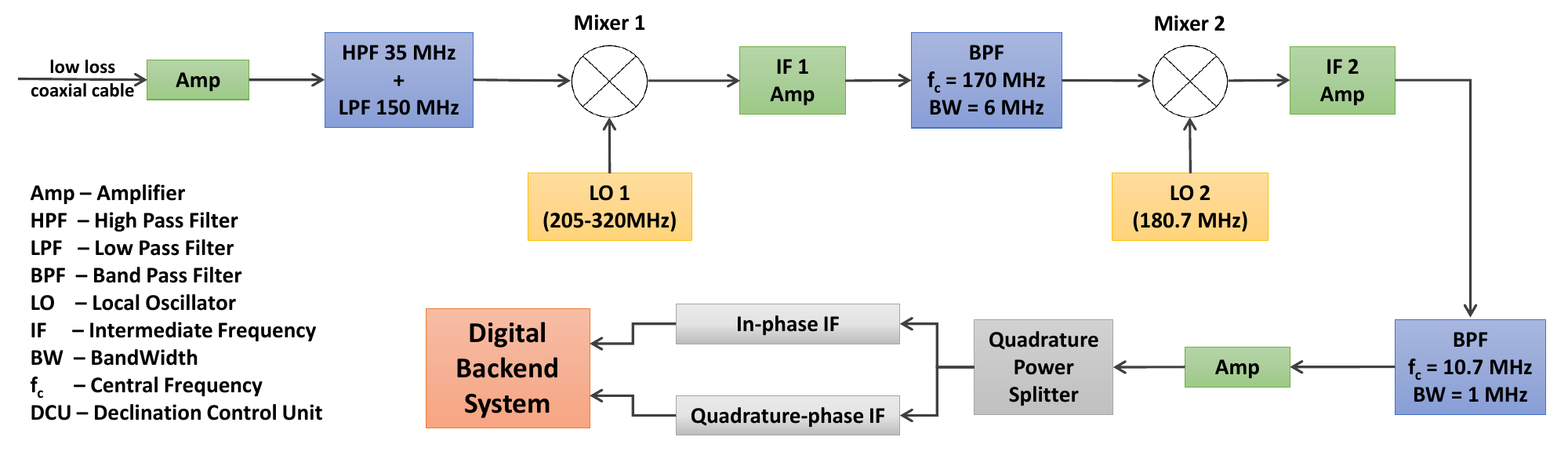}
\caption{Block diagram of an analog back-end receiver chain in the CRB.}
\label{fig:Abackend_graph_pre}
\end{figure}

The configuration and front-end receiver schematic of the GRAPH EW and NS groups are shown in Figure \ref{fig:Afrontend_graph_pre}.
In each EW group, the signal from an LPDA is first passed through a High-Pass Filter (HPF) with a 35\,MHz cutoff, then amplified by $\approx$\,30\,dB using a low-noise amplifier (Amp), which has $\approx$\,3\,dB noise figure. Using low-loss radio frequency (RF) co-axial cables, the signals from the eight antennas are transmitted to the kiosk, where they are combined using a 8-way power combiner. The combined signal, called the group signal, is then amplified by a group amplifier with a gain of $\approx$\,35\,dB and transmitted to the central receiver building (CRB) via low-loss RF co-axial cables. In the NS antenna groups, after first-stage amplification and filtering, the signals from the four LPDAs pass through a two-stage DCU system which helps steering the antenna group beam response to cover any particular declination/source in the declination range of ${\approx}$\ang{-31}\,S and ${\approx}$\ang{59}\,N. The local zenith is ${\approx}$\ang{14}\,N. Section \ref{sec:dcb} describes the declination control system in detail. In the CRB, the signals from all the groups are amplified and filtered again. Then, using the super-heterodyne technique, the desired frequency signal (i.e., the frequency intended for imaging) in the operating band is down-converted to 10.7\,MHz intermediate frequency (IF) using two stages of RF mixing circuits (see Figure \ref{fig:Abackend_graph_pre}). In the first mixing stage, the signal is up-converted to a first IF of 170\,MHz with the first local oscillator (LO 1) chosen between 205 and 320 MHz, and then passed through a band pass filter (BPF) of 6\,MHz bandwidth. Later in the second mixing stage, the signal is down-converted to a final IF of 10.7\,MHz with the second local oscillator (LO 2) of 170\,MHz and passed through a BPF of 1\,MHz bandwidth. This mixing technique will help in avoiding image frequencies to fall at final IF. After last amplification, the 10.7\,MHz IF signal is split into in-phase and quadrature-phase components (Figure \ref{fig:Abackend_graph_pre}).

\begin{figure}
	\centering
	\includegraphics[width=0.5\textwidth]{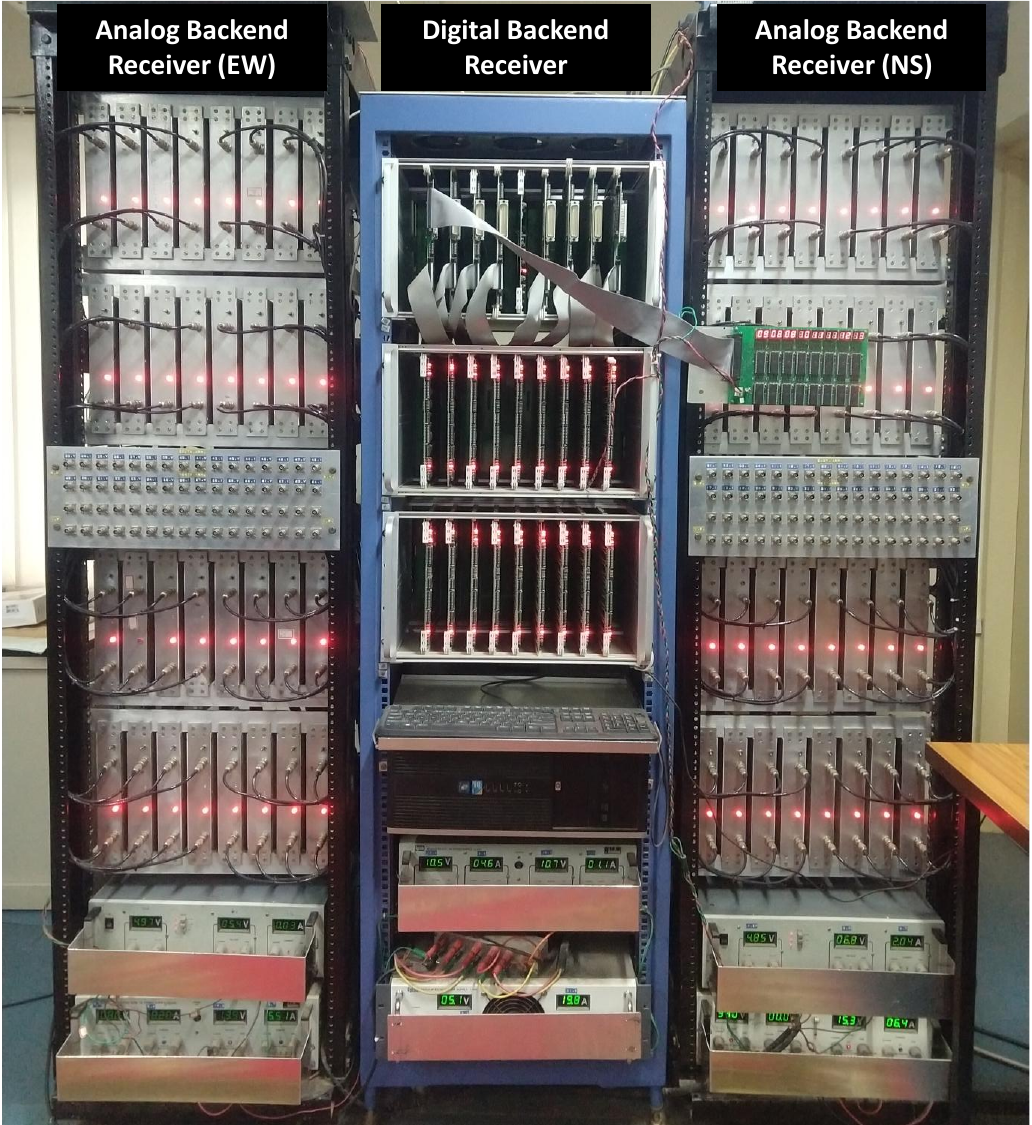}
	\caption{Photograph showing the analog (left \& right racks) and digital (middle rack) back-end receiver systems in the CRB.}
	\label{fig:Dbackend_graph_pre}
\end{figure}
The 64 pairs of in-phase and quadrature-phase signals are transmitted to a digital back-end receiver kept in the central rack (Figure \ref{fig:Dbackend_graph_pre}), which samples the input IF signals at 4\,MHz rate and quantize them into two levels (1-bit). The left and right racks (Figure \ref{fig:Dbackend_graph_pre}) house the 32 analog back-end receiver modules each for the EW and NS groups, respectively. Since the lengths of the low-loss RF cables that transmit the group signals from the kiosks to the CRB are different, the signals from the different groups do not arrive at the same time at the CRB (instrumental delay). Also, there will be differences in the signal arrival times, at the CRB, attributed to the right ascension (RA) and Declination (Dec) of a celestial source (geometrical delay). The above delays are digitally compensated in units of the sampling clock using a digital delay system. The error in delay compensation is 125\,nsec. The digitized signals are then transmitted to a 4096-channel correlator unit, where signals from each antenna group are correlated with the other to obtain the interferometric visibilities. The correlation is performed with Nobeyama correlator chips, and the outputs are integrated for 256\,msec \citep{Ramesh_2006_corr}. The observed visibilities are recorded using an acquisition computer. After pre-processing with custom generated pipelines, the observed data are written into UVFITS format compatible with the Astronomical Image Processing System (AIPS)\footnote{\url{https://www.aips.nrao.edu/}}. As the visibilities are obtained by correlating similarly oriented (\ang{90}) LPDA signals, they are associated to Stokes-I emission (see Section \ref{sec:theory_back}). 
Finally, the UVFITS data are processed using AIPS, and two-dimensional images are obtained.

\section{GRAPH Augmentation} \label{sec:graph_augm}
\subsection{Theoretical Background} \label{sec:theory_back}
\begin{figure}
\centering
\includegraphics[width=0.7\textwidth]{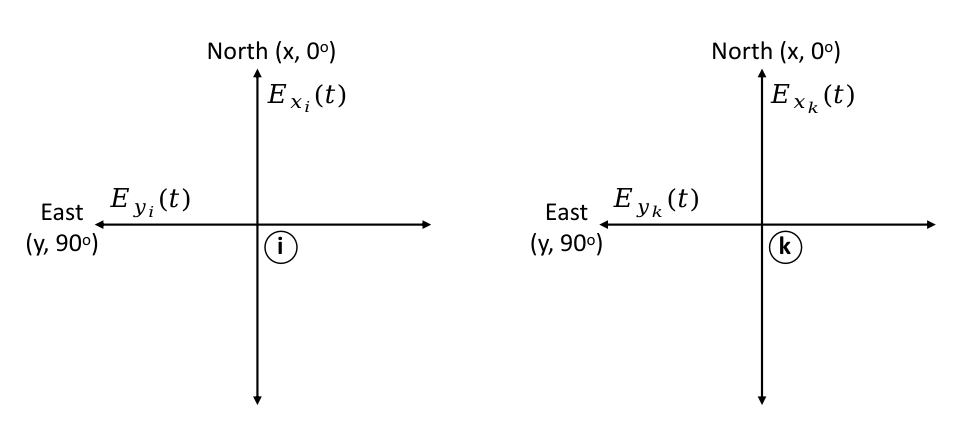}
\caption{Schematic for measuring polarization using orthogonally oriented linearly polarized electric fields.}
\label{fig:nooa}
\end{figure}

Let $E(t)$ be the complex electric field at an observing point at time `t' with $E_x(t)$ and $E_y(t)$ being its orthogonal components. The $x$ axis is along the North (\ang{0} orientation) and $y$ axis is along the East (\ang{90} orientation) as per IAU\citealp{IAU_1973} convention \citep{Weiler1973}. Let $E_{x_i}(t)$, $E_{y_i}(t)$, and $E_{x_k}(t)$, $E_{y_k}(t)$ are the orthogonal electric field components at two observing points \circled{i} and \circled{k}, respectively (Figure \ref{fig:nooa}).
As shown in \cite{Thompson2001}, one can write four possible correlations ($C$) or coherence functions using the electric field vectors measured at the two points in terms of time averaged field intensities and four Stokes visibilities [Stokes I ($I_\nu$), Q ($Q_\nu$), U ($U_\nu$), and V ($V_\nu$)] as given in Equations (\ref{eq:1a})\,-\,(\ref{eq:1d}).
\begin{subequations}\label{eq:1}
	\begin{equation}\label{eq:1a}
		C_{x_i x_k} = \langle E_{x_i}(t) E_{x_k}^*(t) \rangle = \frac{I_\nu + Q_\nu}{2}
	\end{equation}
	\begin{equation}\label{eq:1b}
		C_{y_i y_k} = \langle E_{y_i}(t) E_{y_k}^*(t) \rangle = \frac{I_\nu - Q_\nu}{2}
	\end{equation}
	\begin{equation}\label{eq:1c}
		C_{x_i y_k} = \langle E_{x_i}(t) E_{y_k}^*(t) \rangle = \frac{U_\nu + jV_\nu}{2}
	\end{equation}
	\begin{equation}\label{eq:1d}
		C_{y_i x_k} = \langle E_{y_i}(t) E_{x_k}^*(t) \rangle = \frac{U_\nu - jV_\nu}{2}
	\end{equation}
\end{subequations}

\noindent Reports indicate that if there is any linear polarization originating in the solar corona, it will be obliterated when observed from Earth over a finite bandwidth (${\sim}$\,MHz) due to Faraday rotation in the Solar corona and Earth's ionosphere \citep{Cohen1958, Grognard_McLean_1973, Boischot_Lecacheux_1975, McCauley_2019}. There are a few reports of detection of linearly polarized emission from the solar corona at frequencies $\lesssim$\,100\,MHz, but they are limited to coronal type III radio burst activities, and particularly for narrow bandwidths of the order of 100\,Hz and 10\,kHz \citep{Bhonsle1964,Chin1971}. The circular polarization, unlike the linear component, is unchanged by the Faraday rotation and depolarization, and, therefore can be used as a potential tool to probe the solar corona. Based on the aforementioned observational results, we assume that linear polarization from the solar corona at frequencies 
$\lesssim$\,100\,MHz is negligible, given the 1\,MHz observing bandwidth in our case. Therefore, Stokes Q \& U from the solar corona can be considered negligible for our science cases, and two of the four correlations mentioned above are sufficient to obtain the two Stokes visibilities $I_\nu$ and $V_\nu$, as shown with Equations \ref{eq:2}(a)\,-\,\ref{eq:2}(d).
\begin{subequations}\label{eq:2}
	\begin{equation}\label{eq:2a}
		C_{x_i x_k} = \langle E_{x_i}(t) E_{x_k}^*(t) \rangle = \frac{I_\nu + \cancel{Q_\nu}}{2}
	\end{equation}
	\begin{equation}\label{eq:2b}
		C_{y_i y_k} = \langle E_{y_i}(t) E_{y_k}^*(t) \rangle = \frac{I_\nu - \cancel{Q_\nu}}{2}
	\end{equation}
	\begin{equation}\label{eq:2c}
		C_{x_i y_k} = \langle E_{x_i}(t) E_{y_k}^*(t) \rangle = \frac{\cancel{U_\nu} + jV_\nu}{2}
	\end{equation}
	\begin{equation}\label{eq:2d}
		C_{y_i x_k} = \langle E_{y_i}(t) E_{x_k}^*(t) \rangle = \frac{\cancel{U_\nu} - jV_\nu}{2}
	\end{equation}
\end{subequations}

\noindent Therefore, to observe the circularly polarized radio emission from the solar corona with GRAPH, we decided to install a new set of similar 128 LPDAs with \ang{0} orientation along the existing 128 LPDAs in the NS arm of the array, with uniform gap between the two sets of LPDAs at each location. With this, the total number of LPDAs in augmented GRAPH is 512. The cross-correlations of the signal pairs between \ang{90} oriented antenna groups of EW and NS arms will correspond to Stokes I visibilities (see Equation \ref{eq:2b}). Similarly, cross-correlations of the signal pairs between \ang{90} oriented antenna groups of EW arm and \ang{0} oriented antenna groups of NS arm will correspond to Stokes V visibilities (see Equation \ref{eq:2c}). Effectively, 384 LPDAs (256 in the EW arm with the old 128 in the NS arm) will be involved in Stokes I measurements as earlier, and similarly, 384 LPDAs (256 in the EW arm with the new 128 in the NS arm) will be involved in Stokes V measurements. The DCP of the quiet/undisturbed corona is expected to vary from $\sim$\,3\% to $\lesssim$\,0.1\% in the 40\,-\,150 MHz range \citep{Sastry_2009ApJ}. Therefore, the new system should have high sensitivity to detect weakly polarized signals. Previous studies show that the thermal flux density of the solar corona in the 40\,-\,150 MHz range can vary from $\sim$\,600 to 80,000 Jy \citep{smerd1950,sheridan1985}. From the aforementioned DCP, one can expect a circularly polarized flux of $\approx$\,20\,-\,80\,Jy. Using the various parameters listed in Table \ref{tab:GRAPH} and the typical average brightness temperatures of the background sky (9500\,K and 460\,K at 40\,MHz and 150\,MHz, respectively; \citealp{Haslam_1982,deoliveira_2008}), we find that the theoretical sensitivity of GRAPH can be in the range of ${\approx}$\,2\,-\,3\,Jy in the above frequency range, for 384 antennas, 1\,MHz integration bandwidth, and 10\,sec integration time.

\subsection{Front-end Receivers near the antennas}
\label{sec:frontend}
\begin{figure}
\centering
\includegraphics[width=\textwidth]{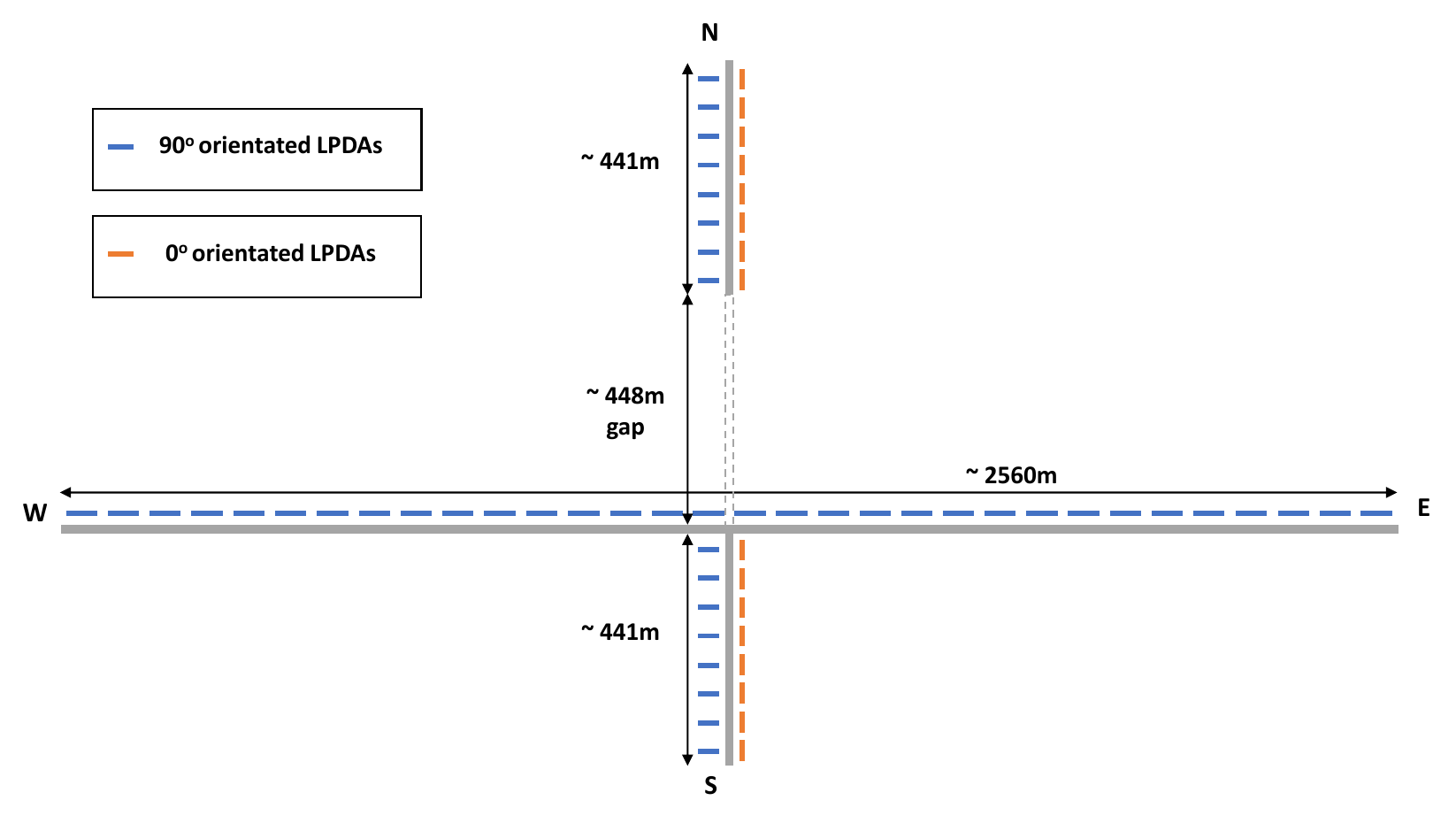}
\caption{Layout of the augmented GRAPH array (not to scale), after augmentation. Each blue line in the EW arm represents group of 8 LPDAs in y-orientation (\ang{90} orientation w.r.t the celestial north). In the NS arm, each blue line  represents groups of 8 LPDAs in y-orientation, and each red line represent groups of 8 LPDAs in x-orientation (\ang{0} orientation w.r.t the celestial north).}
\label{fig:layout_aug}
\end{figure}

\begin{figure}
\centering
\includegraphics[width=0.8\textwidth]{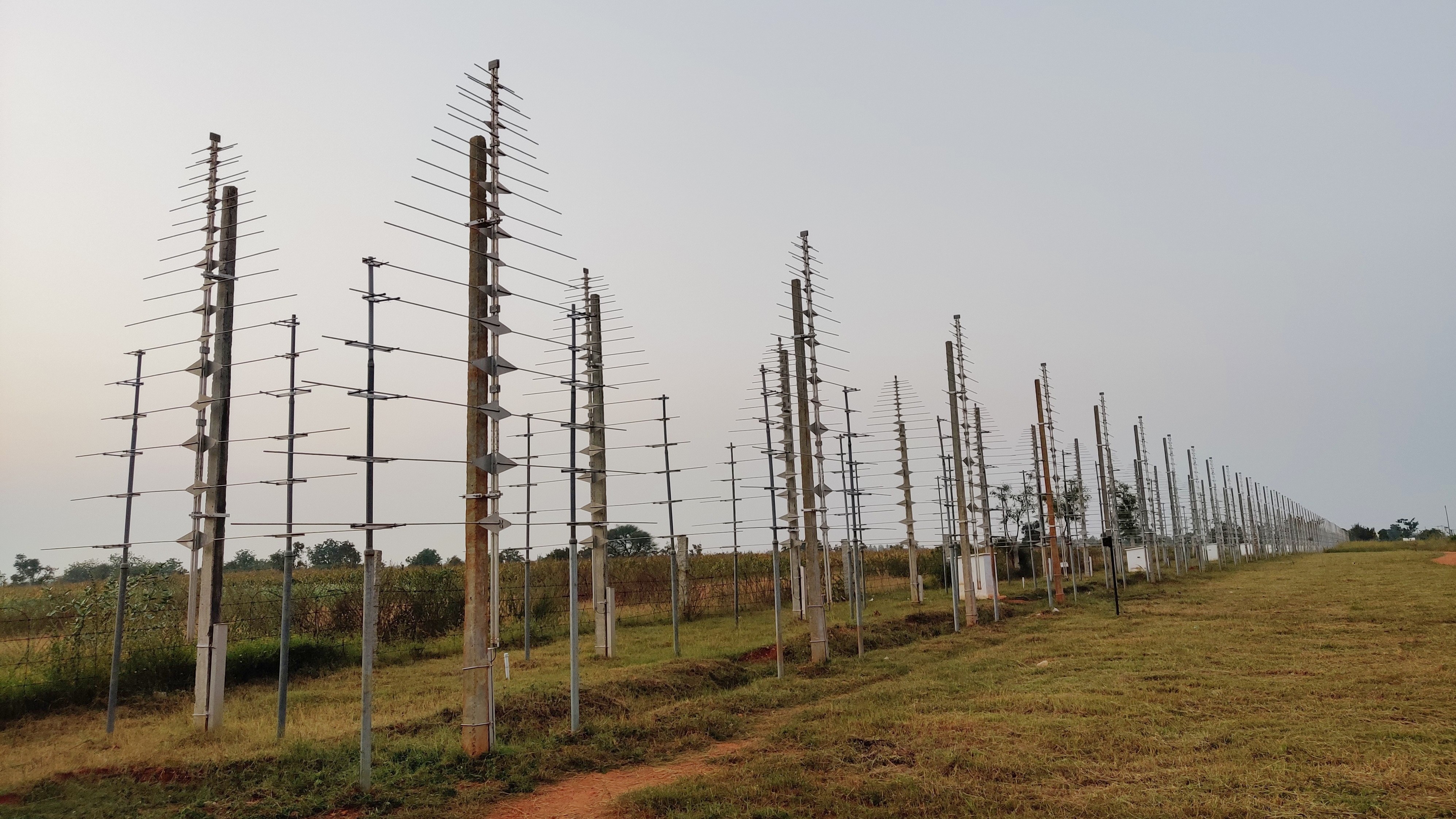}
\caption{Photograph showing a section of the South arm in the augmented GRAPH. The newly installed LPDAs (\ang{0} orientation) are in the foreground, while the old LPDAs (\ang{90} orientation of GRAPH array) are in the background.}
\label{fig:southarm_aug}
\end{figure}

As mentioned in the previous section, 128 new LPDAs were fabricated and installed, with \ang{0} orientation, adjacent to the existing GRAPH antennas in the NS arm, as shown in the schematic in Figure \ref{fig:layout_aug}.
The separation between the two antenna lanes is 4\,m. 
The mutual coupling between the two orthogonal antennas with a 4\,m spacing is 
$\lesssim$\,--30 dB at 40\,MHz, and $\lesssim$\,--50 dB at 150\,MHz \citep{Balanis_2016}. Since the coupling is weak, it may not alter the DCP error due to the intrinsic polarization leakage of an LPDA \citep{Sayuf_2026}, which is accounted for in the calibration process. A section of the south arm of the augmented GRAPH array is shown in Figure \ref{fig:southarm_aug}.
Necessary AC/DC biasing systems, DC cables, RF cables, and analog modules, such as low-noise amplifiers and filters, for the new antennas were characterized and installed in the field. For the augmented GRAPH, eight antennas in the NS arm are combined into a group (both in the old and new set of 128 LPDAs), doubling the collecting area of an NS group compared to the earlier configuration of the GRAPH (Figure \ref{fig:southarm_pre}). 

So, there are now eight LPDAs per each group in augmented GRAPH. The total number of groups is 32 in the EW arm (with \ang{90} oriented LPDAs), 16 in the NS arm with \ang{90} oriented LPDAs, and another 16 in the NS arm with \ang{0} oriented LPDAs (Figure \ref{fig:layout_aug}). Some parameters in Table \ref{tab:GRAPH} have undergone a change due to the augmentation; those parameters alone are given in Table \ref{tab:GRAPHIC}.
\begin{table}[h]
	\caption{Changes to Table \ref{tab:GRAPH} parameters after GRAPH augmentation.}
	\label{tab:GRAPHIC}
	\begin{tabular}{ll}
		\hline
		\textbf{Parameter} & \textbf{Value} \\
		\hline
		Number of antenna groups & 64 [32 EW \& 32 NS (16 each for \ang{0} \& \ang{90} orientations)] \\
		LPDAs per group & 8 (EW) \& 8 (NS) \\
		Total number of LPDAs & 512 (384 in \ang{90} orientation, \& 128 in \ang{0} orientation) \\
		Effective collecting area & 192\,$\rm \lambda^2$ each for Stokes I \& Stokes V\\
		\hline
	\end{tabular}
\end{table}

\begin{figure}
\centering
\includegraphics[width=0.9\textwidth]{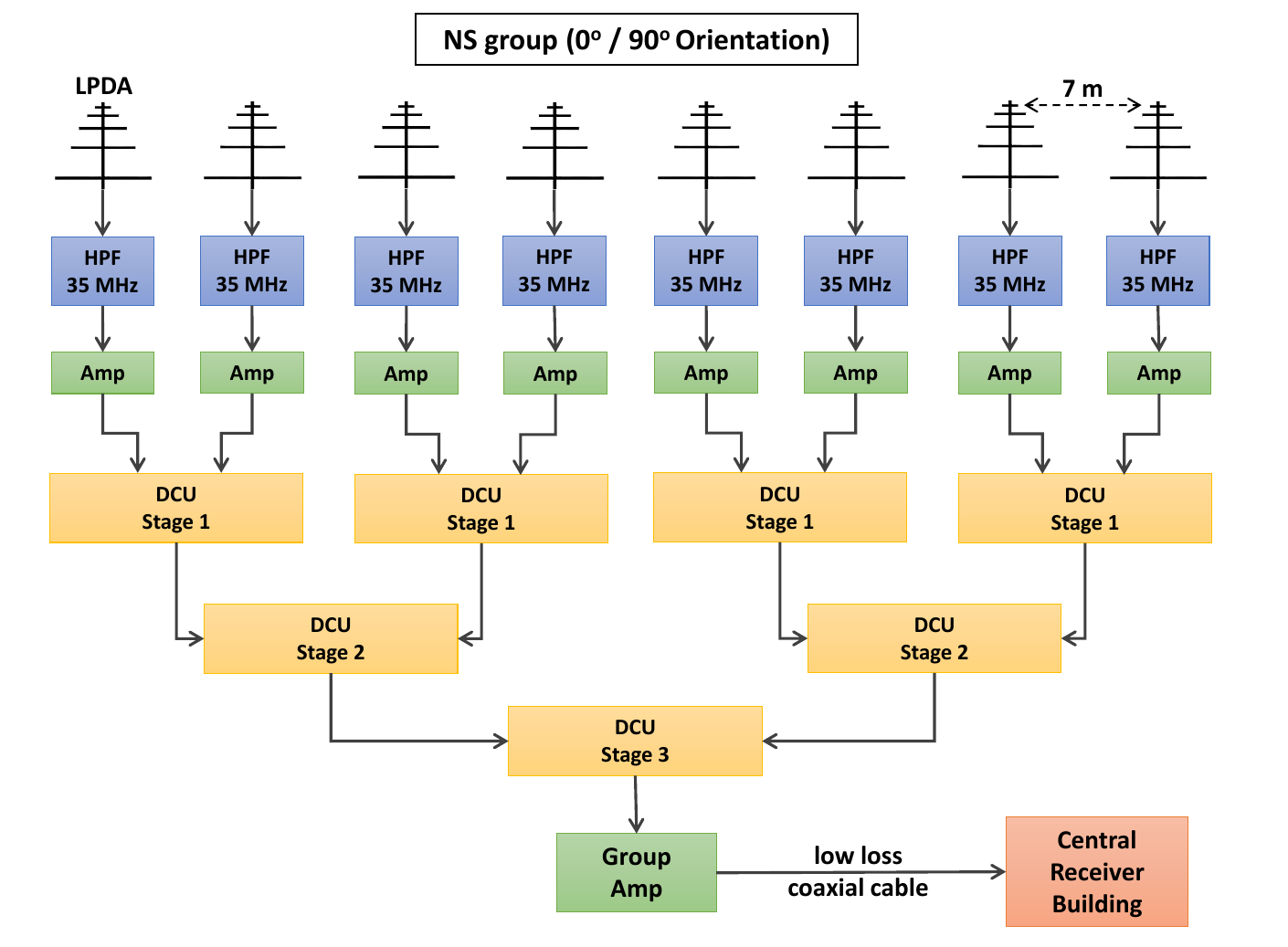}
\caption{Block diagram of an analog front-end receiver system in an NS group of the GRAPH, after augmentation.}
\label{fig:Afrontend_graph_aug}
\end{figure}

\begin{figure}
\centering
\begin{minipage}{\textwidth}
\centering
\includegraphics[width=0.95\linewidth]{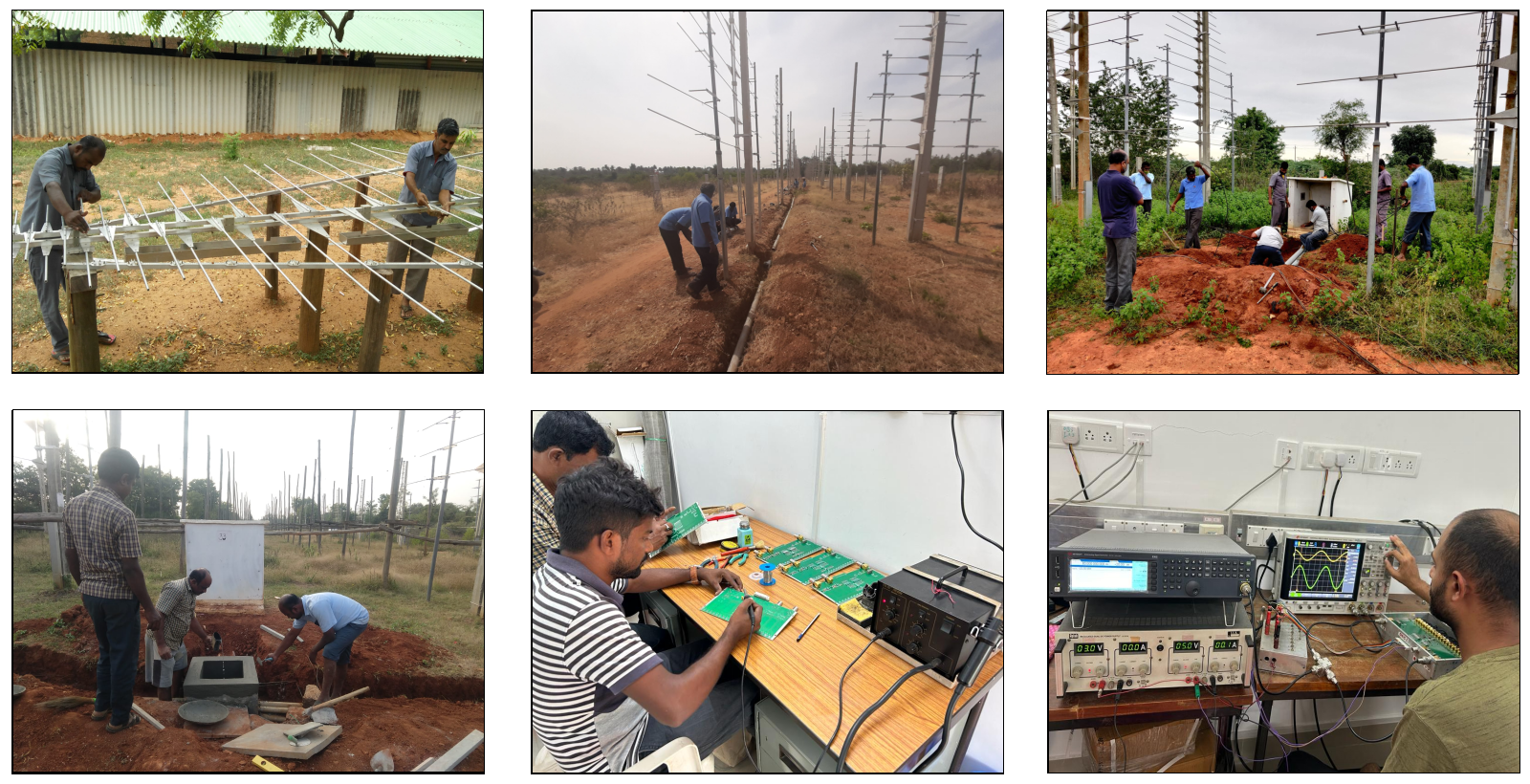}
\end{minipage}
\begin{minipage}{\textwidth}
\centering
\includegraphics[width=0.95\linewidth]{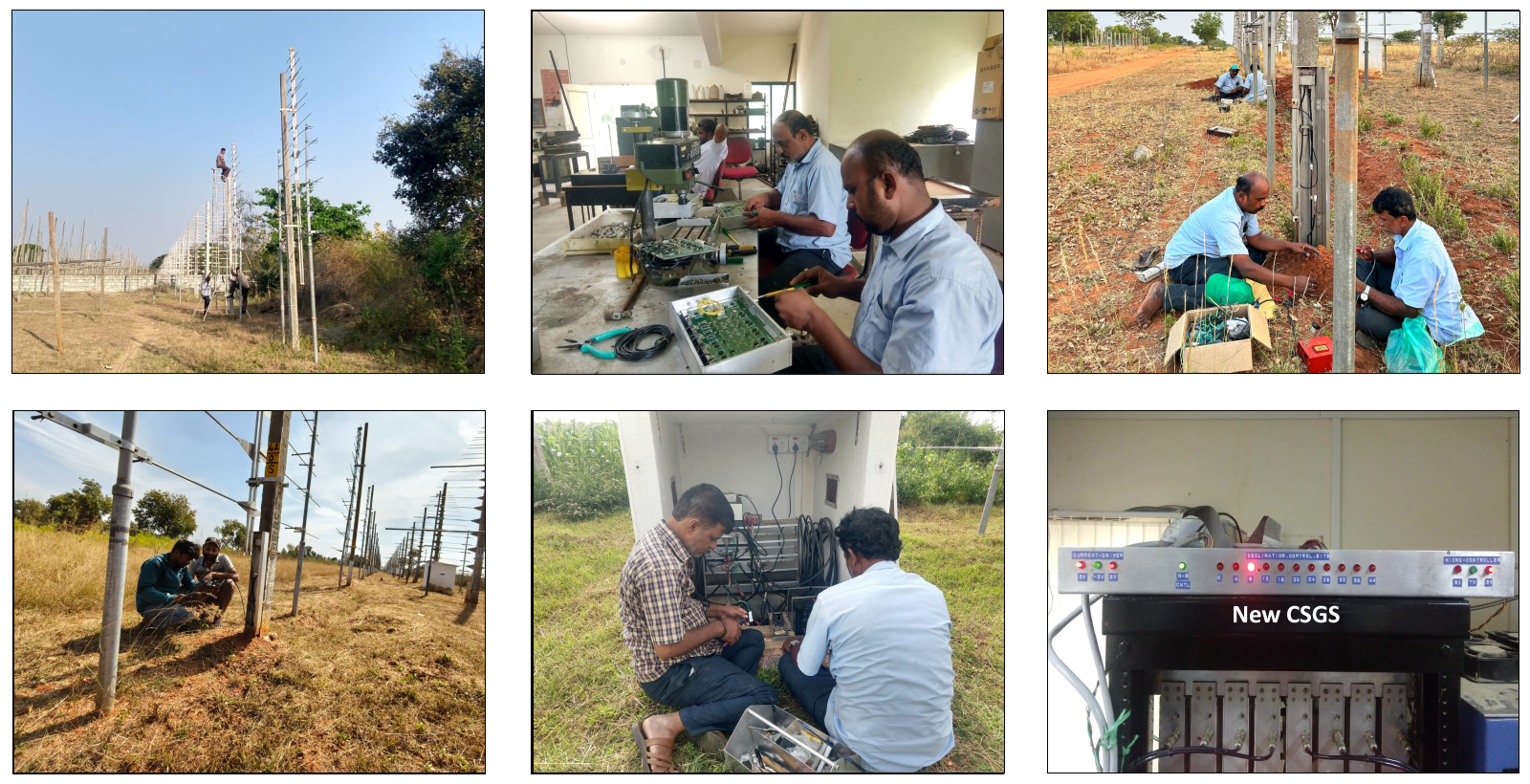}
\end{minipage}
\caption{Photograph collection showing various stages of the GRAPH augmentation activities.}
\label{fig:aug_works}
\end{figure}

The configuration and front-end receiver system of a NS group in the augmented GRAPH array are shown in Figure \ref{fig:Afrontend_graph_aug}.
To steer the response of an antenna group in the NS arm to a desired declination, the signals from the individual antennas are appropriately delayed using a 3 stages of DCUs. As the group size is doubled in the NS arm, the group beam width reduced by approximately half the width and hence the beam should be steered in the smaller steps of \ang{4} instead of \ang{8} in the case of before augmentation. So new DCUs are designed and developed to meet this requirement and further detailed related to DCUs and their control system are discussed in the Section \ref{sec:dcb}.
Figure \ref{fig:aug_works} shows the various work carried out in the lab as well as near the antennas in the field by the Gauribidanur observatory team during the augmentation activity.
\noindent

\subsection{Back-end receivers at the CRB}\label{sec:backend}

In the CRB, the analog and digital back-end receivers of the GRAPH system before augmentation are  used without any modification, since the total number of antenna groups in the array remains unchanged at 64 in the augmneted GRAPH also. However, the RF and IF ports of the \ang{0} and \ang{90} orientations were separated for easy monitoring. The software for segregating the $C_{y_iy_k}$ and $C_{x_iy_k}$ correlations were developed and tested. All the pre-processing pipelines and the software programs which convert the raw correlation output into AIPS compatible UVFITS files were also developed and tested. Currently, for each observation, two UVFITS files are generated: one for the $C_{y_iy_k}$ correlation data and another for the $C_{x_iy_k}$ data. Due to 4\,m separation between the \ang{0} and \ang{90} antennas in the NS arm, there would be a slight systematic offset in the UV-coordinate location and change in UV coverage. The effects of these issues were accounted by synchronizing their phase centers and aligning the coordinates in software.

\subsection{DCU and Control Signal Generation System}\label{sec:dcb}
\begin{figure}
\centering
\includegraphics[width=\linewidth]{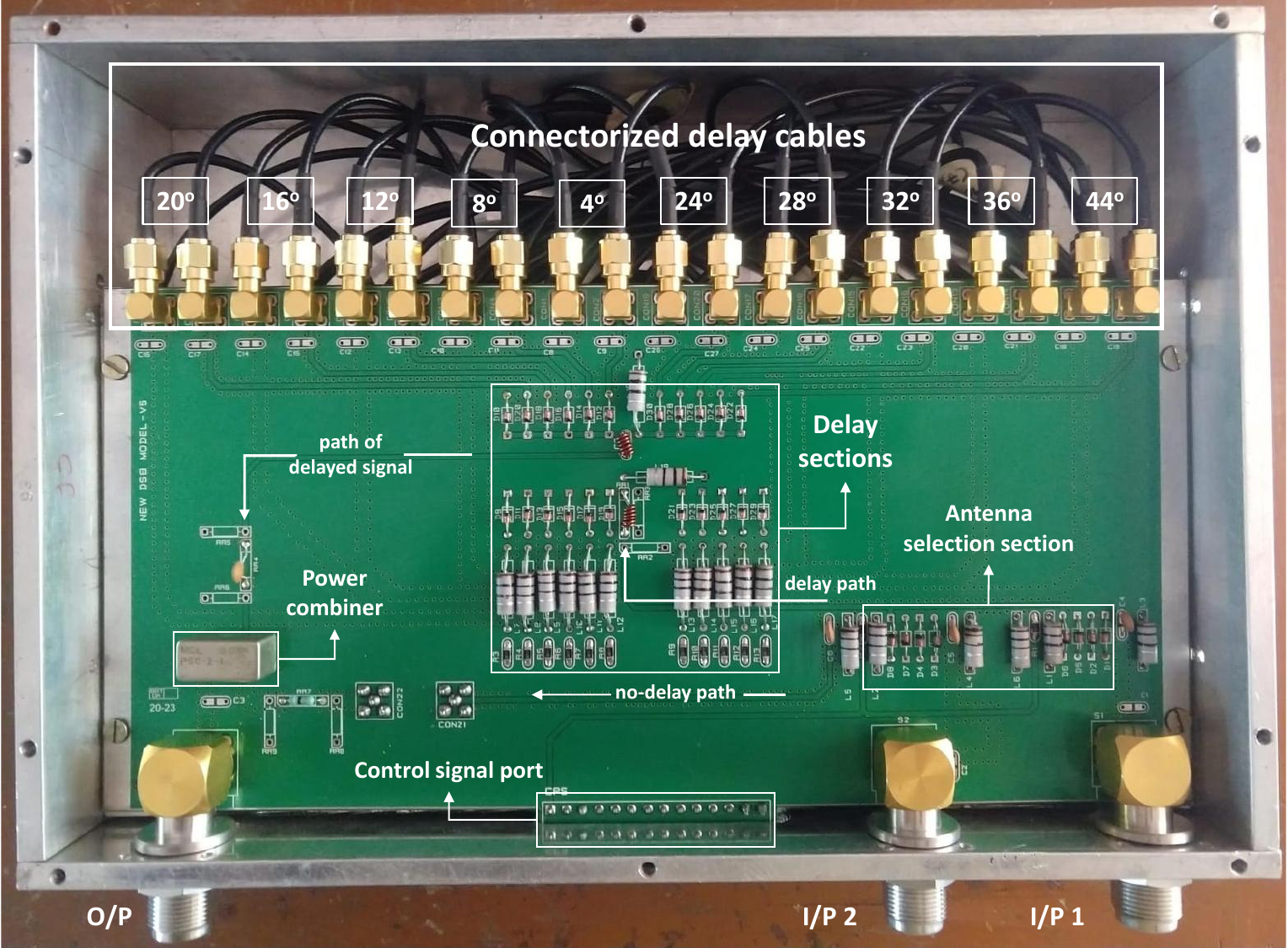}
\caption{Photograph of the DCU in the augmented GRAPH. The angles mentioned on the cables near the top of the PCB correspond to different zenith angles by which the group beam can be tilted.}
\label{fig:dcbs}
\end{figure}

As mentioned earlier, a DCU is used to steer the electrical response of a group of antennas \citep{Landecker1984,Tingay_2013,vanhaarlem_2013,Warnick_2018}, i.e., the group beam, electronically along the declination axis. A DCU is RF module with a printed circuit board (PCB), RF components, and cables inside. The PCB is populated with electronic components such as diodes, capacitors, inductors, power combiners, and others. The top view of the new DCU, designed, developed, and characterized for the augmented GRAPH array, is shown in Figure \ref{fig:dcbs}.
The board features two input terminals (I/P\,1 \& I/P\,2) and one output terminal (O/P); the input terminals receive signals from two individual LPDAs. A key component of the DCU is a initial switch called antenna selection section which can alter the propagation paths of the input radio signals. It is controlled by a diode circuit. This routing system consists of two paths (no-delay path and delay path) : one transmits the radio signals as they are, i.e., without any delay, while the other introduces a time delay with some attenuation respectively. Furthermore, the delay path comprises eleven subsections, which are also controlled by the diode switches. Except for the zero-delay (\ang{0} tilt) subsection, all other 10 subsections carry coaxial cables of different lengths which can delay and attenuate the signals differently. These eleven subsections are designed with appropriate delay cables such that they provide tilt angles of \ang{0} to \ang{44} in the steps of \ang{4} (tilt angle \ang{40} is skipped). When a DCU receives signals from two antennas, it routes one signal through the no-delay path (depending on tilting the beam towards north or south of the local zenith) and the other one towards delay section where it passes through one of the delay subsections (depending on the required tilt angle), thereby determining the tilt angle for the combined beam of the two antennas. Finally, the two signals are combined using an RF power combiner and transmitted from the DCU's output terminal to the subsequent section in the receiver chain.

Since the augmented GRAPH array consists of eight antennas in each group in the NS arm, they are organized into a 3-stage Christmas-tree configuration (Figure \ref{fig:Afrontend_graph_aug}); the group beam is steered toward the desired Declination using first-, second-, and third-stage DCUs. 
The ten subsections of the DCU from \ang{0} tilt angle to \ang{36} are used to observe sources in the declination range $\approx$ \ang{-24.3} to \ang{51.7} which include Sun and other major low-frequency calibrators like Cygnus-A, Taurus-A, Virgo-A, etc. The \ang{44} delay subsection enables the observation of Cassiopeia-A (declination $\approx$\,\ang{59}) in northern direction and Sagittarius-A (declination $\approx$\,\ang{-29}) in southern direction. The minimum and maximum delay achievable with a 3-stage DCU module for \ang{4} and \ang{44} are 1.63\,nsec and 113.4\,nsec, respectively. Another notable feature of the DCU is its ability to steer the composite beam along the North and South declinations symmetrically.

\begin{figure}
\centering
\includegraphics[width=0.9\textwidth]{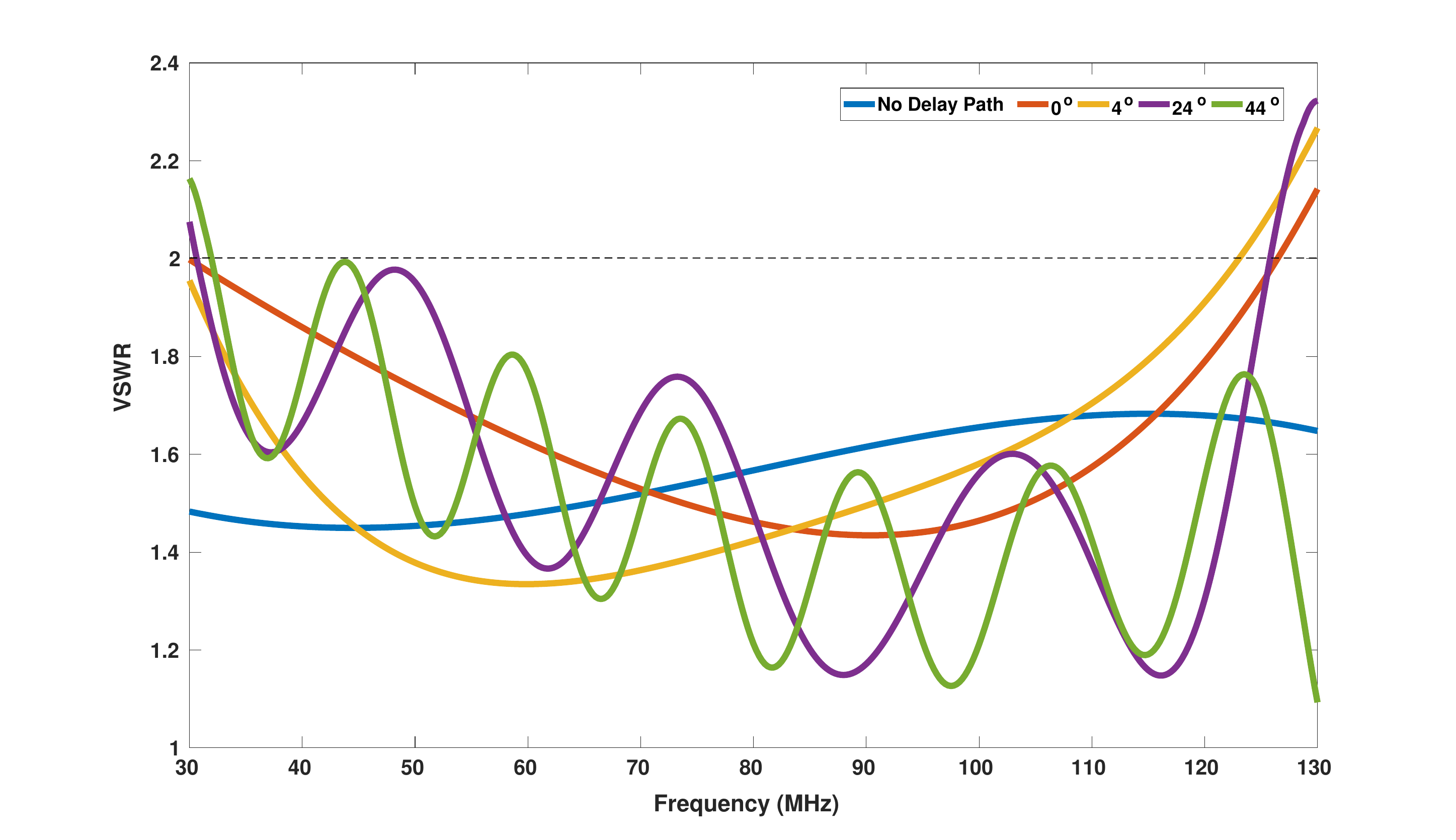}
\caption{VSWR of a Stage 2 DCU for a few declination angles (or delay paths).}
\label{fig:dcb_s11_stg2}
\end{figure}

\begin{figure}
\centering
\includegraphics[width=0.9\textwidth]{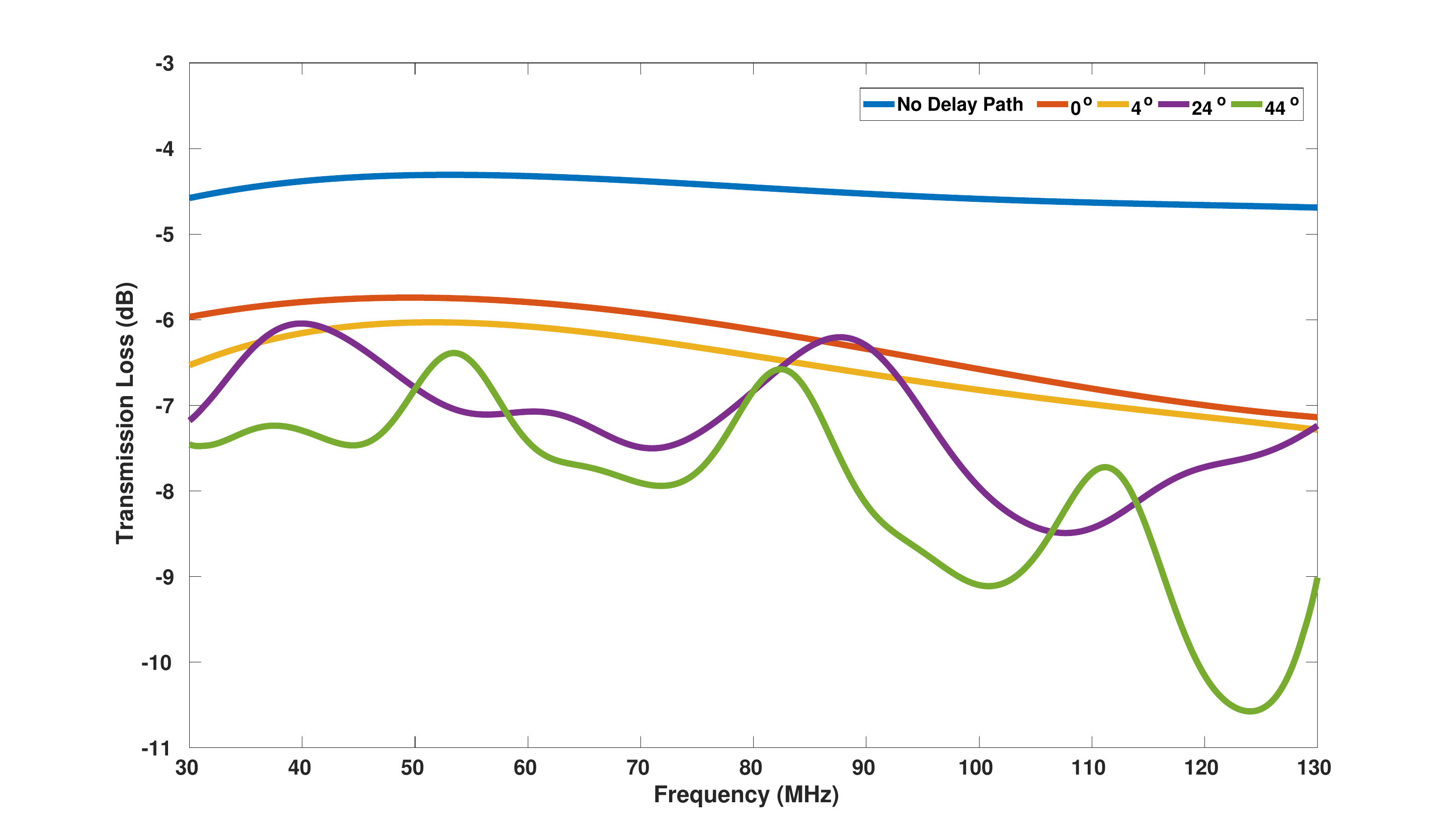}
\caption{Transmission loss of a Stage 2 DCU for a few declination angles (or delay paths).}
\label{fig:dcb_s21_stg2}
\end{figure}

\begin{figure}
\centering
\includegraphics[width=0.9\textwidth]
{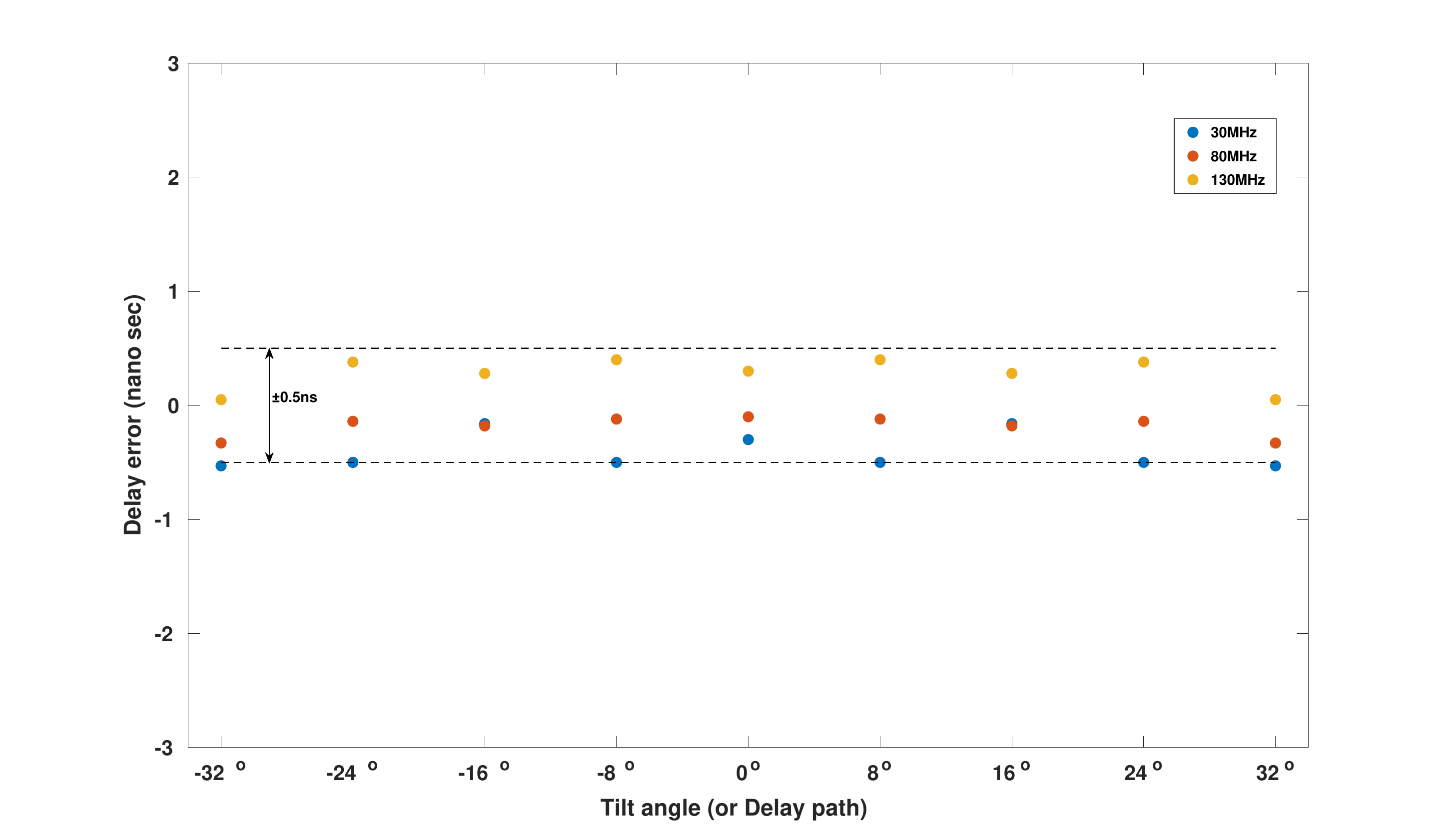}
\caption{Comparison of delay errors for different tilt angles (or delay paths) for the GRAPH Stage 2 DCU.}
\label{fig:dcb_oldnew_dly2}
\end{figure}

The DCU has been configured with minimal reflection loss. Figure \ref{fig:dcb_s11_stg2} shows the VSWR of a DCU used in the second stage (Figure \ref{fig:Afrontend_graph_aug}).
The blue, orange, green, yellow, and black profiles correspond to the VSWR of the DCU for the no-delay path, and delay paths corresponding to the \ang{0}, \ang{4}, \ang{24}, and \ang{44} tilt angles, respectively. The VSWR values $\lesssim$\,2.2 for all paths indicate that the impedance of the DCU matches well with the intrinsic system impedance (50\,$\Omega$). Figure \ref{fig:dcb_s21_stg2} shows the transmission loss of the same DCU.
The profile colors mentioned above apply here also. The transmission loss is almost independent of frequency for the no-delay path. However, it increases with frequency and declination angle for all the delay paths. The latter is expected, as the loss increases with tilt angle due to the longer cable lengths required for larger tilt angles. 
We have shown the delay error with frequency in Figure \ref{fig:dcb_oldnew_dly2}. The error is $\pm$0.5\,nsec, which corresponds to a tilt error of ${\approx}\,\ang{0.3}$.

In addition to the input and output RF signal ports, the DCU has a bias port. It has 14 control pins: 10 for the delay sections, 1 for the zero-delay section, 1 for the antenna selection section, and 2 for ground connection. The bias port is connected to the Control Signal Generation System (CSGS) via the control signal cables. The CSGS includes a micro-controller, a current-driver circuit, a constant-current power supply, and a 16-core cable. The latter connects the CSGS with all the DCUs in the field. Before commencing an observation, a desktop computer helps CSGS generate a set of control bits based on the source's declination, which are driven to the field to enable/disable various delay/no-delay sections. For the latter, the CSGS biases the DCUs with a positive/negative voltage ($\approx$ +/-3V).

\section{Calibration}\label{sec:calib}
The response of an antenna `i' with orthogonally polarized feeds `x' (\ang{0} orientation) and `y' (\ang{90} orientation) can be factorized into physically distinct components using the Jones matrix $J_{i}$ \citep{cotton_1999}:
\begin{equation}
	J_i = G_i D_i P_i
\end{equation}%
where the gain term ($G_i$) is written as,
\begin{equation}
	G_i = \begin{pmatrix}
		g_{x_i} & 0 \\
		0 & g_{y_i} 
	\end{pmatrix}
	\label{eqn:gain}
\end{equation}%
where $g_{x_i}$ and $g_{y_i}$ are the complex gain factors for the signals received with the two orthogonal feeds. The leakage term ($D_i$) in equation 3 is,
\begin{equation}
	D_i = \begin{pmatrix}
		1 & d_{x_i} \\
		-d_{y_i} & 1
	\end{pmatrix}
	\label{eqn:leak}
\end{equation}%
where $d_{x_i}$ and $d_{y_i}$ are the fractions of one polarization leaking into the other. The parallactic rotation term ($P_i$) in equation 3 is defined as,
\begin{equation}
	P_i = \begin{pmatrix}
		\cos{\chi} & -\sin{\chi} \\
		\sin{\chi} & \cos{\chi}
	\end{pmatrix}
	\label{eqn:paralax}
\end{equation}%
where ${\chi}$ is the parallactic angle, i.e., misalignment of the antenna coordinate system with respect to the source coordinate system. 

If $I_{\nu},\,Q_{\nu},\,U_{\nu},\,\&\,V_{\nu}$ are the Stokes visibilities corresponding to a radio source, then the possible correlations ($C$) in Muller matrix form that can be obtained on an interferometer baseline between the $i^{th}$ and $k^{th}$ antennas with orthogonal polarized feeds are \citep{cotton_1999}:

\begin{equation} \label{eq:muller_matrix}
	\begin{pmatrix}
		C_{x_i x_k}\\ C_{x_i y_k}\\ C_{y_i x_k}\\ C_{y_i y_k}
	\end{pmatrix}
	= \left(J_i \otimes J_k^*\right) \frac{1}{2}
	\begin{pmatrix}
		1 & 1 & 0 & 0\\
		0 & 0 & 1 & j\\
		0 & 0 & 1 & -j\\
		1 & -1 & 0 & 0\\
	\end{pmatrix}
	\begin{pmatrix}
		I_{\nu} \\ Q_{\nu} \\ U_{\nu} \\ V_{\nu}
	\end{pmatrix}
\end{equation}%
where the symbol `$\otimes$' denotes the outer matrix product.
As mentioned in Section 3.1, Stokes Q \& U from Sun (target source) are negligible in our case, and therefore they can be assumed to be zero. We would also like to add here that the GRAPH system is equipped to compute $C_{x_i y_k}$ \& $C_{y_i y_k}$ correlations only among all four correlations mentioned in Section 3.1. So, we can obtain only Stokes I \& V from them. Hence, Equation \ref{eq:muller_matrix} can be rewritten as:

\begin{equation} \label{eq:muller_matrix_half}
	\begin{pmatrix}
		0\\ C_{x_i y_k}\\ 0\\ C_{y_i y_k}
	\end{pmatrix}
	= \left(J_i \otimes J_k^*\right) \frac{1}{2}
	\begin{pmatrix}
		1 & 1 & 0 & 0\\
		0 & 0 & 1 & j\\
		0 & 0 & 1 & -j\\
		1 & -1 & 0 & 0\\
	\end{pmatrix}
	\begin{pmatrix}
		I_{\nu} \\ 0 \\ 0 \\ V_{\nu}
	\end{pmatrix}
\end{equation}%
\noindent The above matrix multiplication reduces to,
\begin{equation}\label{eq:Cyiyk}
	C_{y_i y_k} = \frac{1}{2} g_{y_i}g_{y_k}^* I_{\nu}
\end{equation}
\begin{equation}\label{eq:Cxiyk}
	C_{x_i y_k} = \frac{1}{2} g_{x_i}g_{y_k}^* \left(\left(d_{x_i}-d_{y_k}^*\right) I_{\nu} +jV_{\nu} \right)
\end{equation}

\noindent From Equations \ref{eq:Cyiyk} \& \ref{eq:Cxiyk}, one can infer that 
$C_{y_i y_k}$ has contribution from Stokes I only. $C_{x_i y_k}$ has contribution from Stokes V, and Stokes I leakage. In order to estimate the `true' Stokes I and Stokes V visibilities, the system parameters like complex gains ($g_{x_i}$, $g_{y_i}$, $g_{y_k}$) and polarization leakage $\left(d_{x_i}-d_{y_k}^*\right)$ must be calculated and applied to the observed visibilities (or correlations). We have used observations of `unpolarized' radio point sources (A-team sources, viz., Virgo-A, Taurus-A, Cygnus-A, and Cassiopeia-A) as calibrators to calculate the system parameters (see, e.g. \citealp{Devojyoti_2025}).

\begin{figure}
\centering
\includegraphics[width=0.9\textwidth]{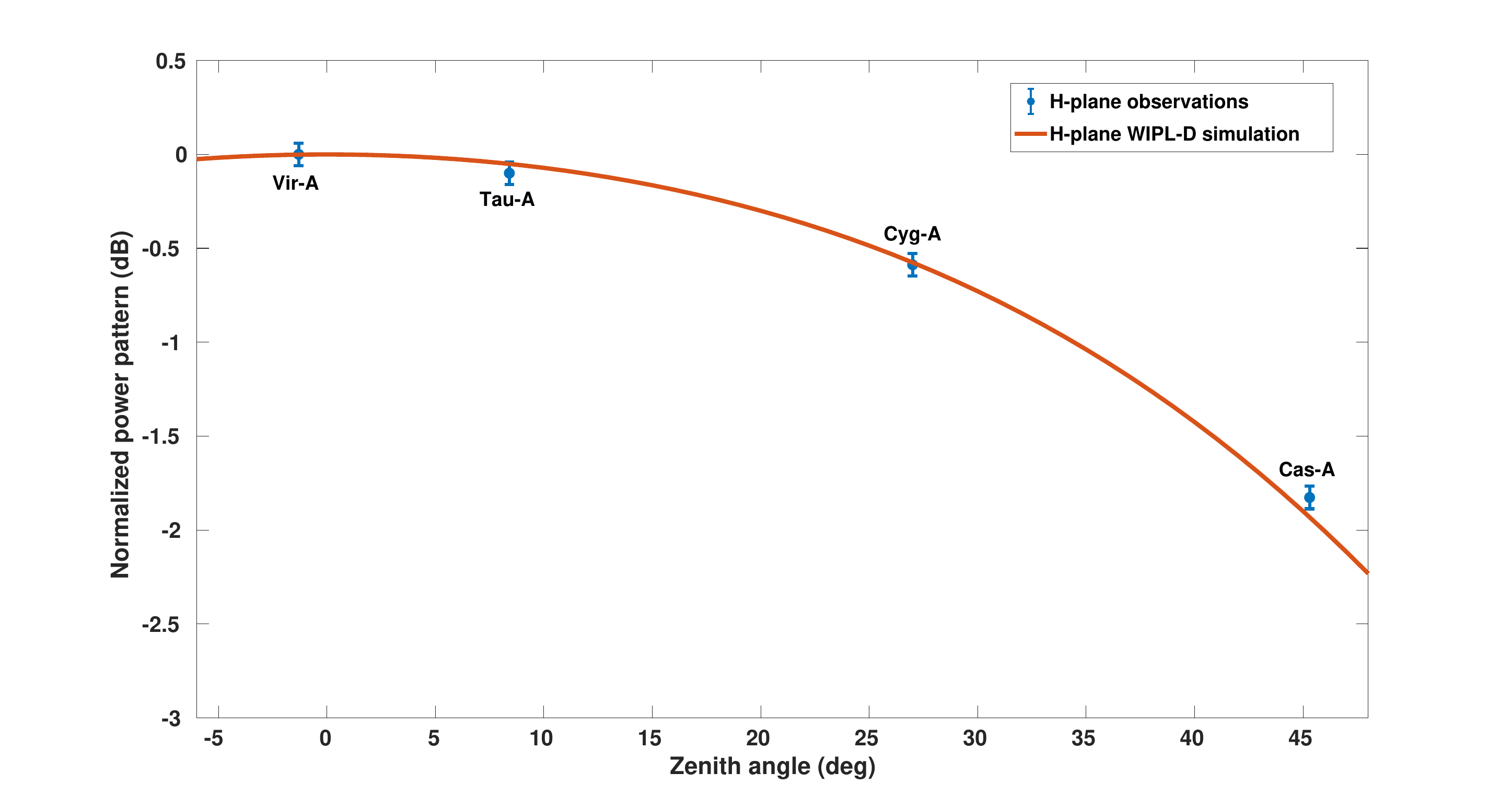}
\caption{Normalized H-plane pattern for the GRAPH LPDA estimated using GRIP observations of main A-team sources at 51 MHz, and the corresponding WIPL-D simulated values. There is a very good correspondence between the simulations and the observations.}
\label{fig:hplane_gain}
\end{figure}

\begin{figure} 
\centering
\includegraphics[width=0.9\textwidth]{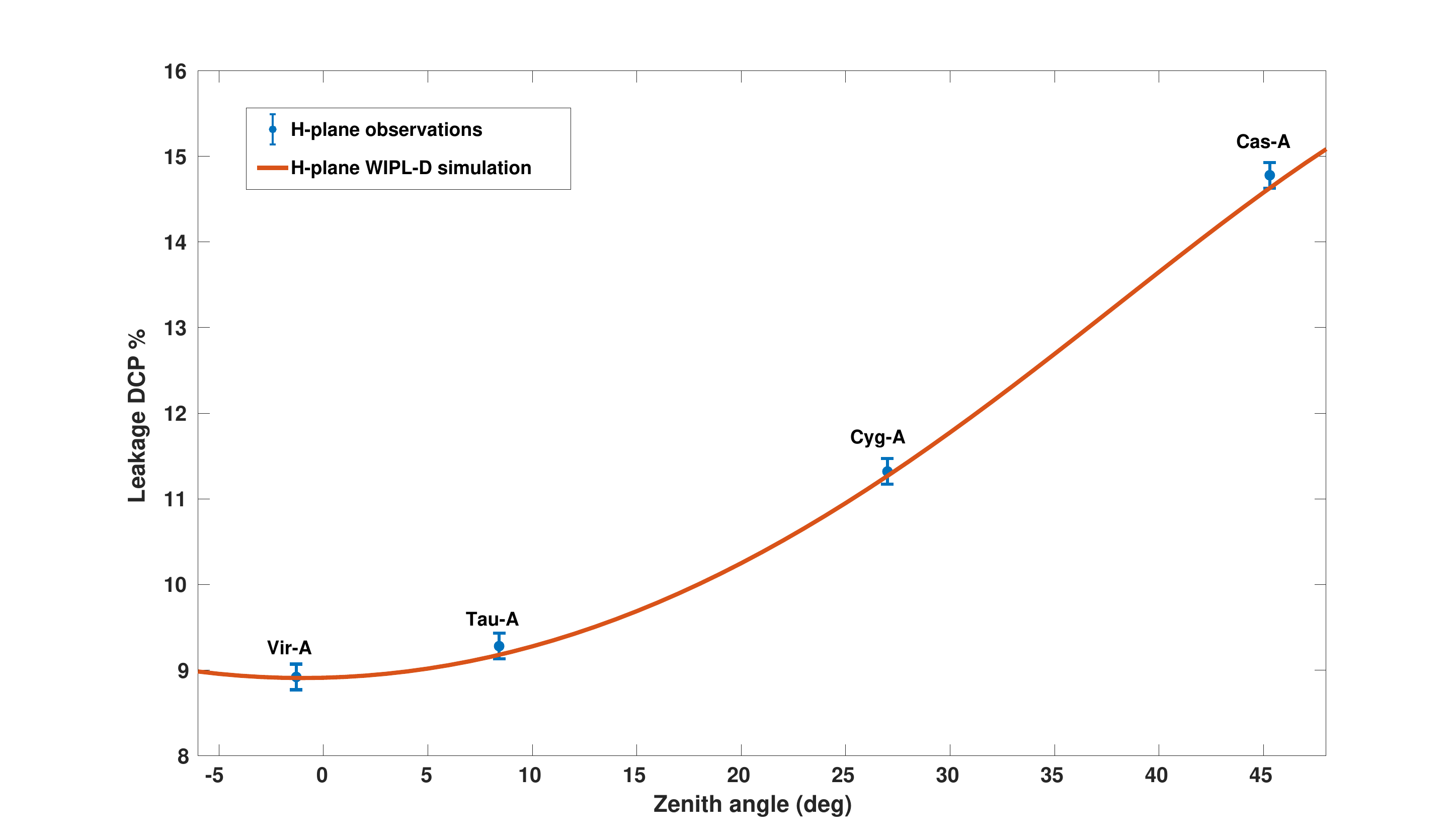}
\caption{System leakage for the GRAPH LPDA estimated using GRIP observations of main A-team sources at 51 MHz, and the corresponding WIPL-D simulated values \citep{Sayuf_2026}. There is a good correspondence between the leakages estimated from the observations and simulations.}
\label{fig:hplane_leak}
\end{figure}

We can also use GRAPH antenna response from either the observations or electromagnetic (EM) simulations, to correct for the antenna dependent parameters. For instance, in order to estimate the zenith angle (declination axis in the present case) based gain and polarization leakage of the GRAPH antenna, we have used observations with Gauribidanur Radio Interferometric Polarimeter (GRIP) in the observatory, since the design of the LPDAs in GRIP and GRAPH are identical \citep{Grip_2008,Sayuf_2026}. We observe various bright sources (A-team) at different declinations during their meridian transit in the Gauribidanur observatory. After calibration, the results are compared with the EM simulation results of the GRAPH antenna structure, developed and simulated using the WIPL-D software package\footnote{\url{https://wipl-d.com/products/wipl-d-pro-cad/}} (v21, WIPL-D, 2024). There is good agreement between the observations and simulation results, indicating that WIPL-D simulations can be used in calibration for antenna primary beam based corrections between the calibrator and target sources. Figures \ref{fig:hplane_gain} and \ref{fig:hplane_leak} show the H-plane gain and polarization leakage, respectively, of the GRAPH LPDA at 51 MHz. To implement the calibration strategy, a cross-platform pipeline using Python \& AIPS was developed. The calibration techniques for Sun's Stokes-I and Stokes-V data are discussed in following Sections \ref{sec:calibsi} and \ref{sec:calibsv} respectively. The subscripts `$cal$' and `$sun$' for various terms in the equations present in these two sections represent the terms associated to calibrator and Sun respectively.

\subsection{Stokes I calibration}\label{sec:calibsi}

As mentioned earlier, in order to calibrate the target source, i.e., the Sun, few unpolarised calibrator sources (typically A-team sources whose $Q_\nu$, $U_\nu$ \& $V_\nu$ are negligible) are observed before/after the Sun's local meridian transit. The combined UVFITS data of the Sun and calibrator, associated to Stokes-I correlation, is loaded into AIPS. Considering a frequency-specific flux model (e.g. \citealp{Perley_2017}) for the calibrator, the observed calibrator data, can be expressed from Equation \ref{eq:Cyiyk} as below:
\begin{equation}\label{eq:Cyiyk_cal}
	C_{y_i y_k, cal} = \frac{1}{2}\;\; g_{y_i, cal}\;\; g_{y_k, cal}^* \;\; I_{\nu, cal\,model}
\end{equation}
\noindent By incorporating the calibrator flux model into AIPS, a solution table (SN) for the calibrator is generated using `CALIB' task. The table consists of complex gain solutions for each y-oriented antenna group which are used to calculate their respective complex gain terms ($g_{y, cal}$). These solutions are direction dependent. So, before applying to the Sun data, they have to be updated for the Sun's sky position. The main factors are the antenna gain and DCUs gain differences between the calibrator and Sun positions. They are obtained from antenna WiPL-D simulations and DCU characterisations, respectively. The SN table from calibrator is exported as a .TXT file and read into Python along with the antenna \& DCU files. The complex gain solutions for Sun's position are updated and exported into AIPS as a new SN table in .TXT format. Using the updated gain solutions for Sun, the calibration table, which has the complex gain terms ($g_{y, sun}$) of the respective antenna groups, is generated and applied to the Sun data to obtain the corrected Stokes-I visibilities in AIPS (Equation \ref{eq:Iv_cor}). The visibility phases alone are processed once through self-calibration to get the corrected complex gain phases and thus the final visibilities. During the analysis, the `bad' antennas and baselines are flagged to fine-tune the calibration and improve the image quality. With the final set of corrected visibilities ($I_{\nu, sun~(corrected)}$), the Sun's Stokes-I image is obtained using the `IMAGR' task in AIPS.

\begin{equation}\label{eq:Iv_cor}
	I_{\nu, sun~(corrected)} = \frac{2\, C_{y_i y_k, sun}}{g_{y_i, sun}\;\; g_{y_k, sun}^*}
\end{equation}

\subsection{Stokes V calibration}\label{sec:calibsv}
To calibrate the Stokes V data corresponding to Sun observations, we have used the Stokes V leakage observed with the unpolarized calibrator sources since there are no strong circularly polarized calibrators at low radio frequencies. As the technique relies on leakage flux, it would be beneficial to observe strong sources like the A-team sources to get better SNR. The observed visibility data in both the UVFITS files related to Stokes-I and Stokes-V are read into Python. 
For an unpolarized calibrator source, Equation \ref{eq:Cxiyk} can be written as below: 
\begin{equation}\label{eq:Cxiyk_cal}
	C_{x_i y_k, cal} = \frac{1}{2}\;\; g_{x_i, cal}\;\; g_{y_k, cal}^*\;\;  \left[\left(d_{x_i, cal}-d_{y_k, cal}^*\right) I_{\nu,cal\,model} \right]
\end{equation}
\noindent Only the complex gain term $g_{x_i, cal}$ is unknown in the above equation 13. In the other terms, $C_{x_i y_k, cal}$ is the observed correlation, $g_{y_k, cal}^*$ is estimated during Stokes-I calibration, $\left(d_{x_i, cal}-d_{y_k, cal}^*\right)$ for calibrator position can be obtained from WiPL-D simulations, and $I_{\nu,cal}$ is known from the calibrator flux model. So, the complex gains of x-oriented antenna groups ($g_{x, cal}$) can be calculated following a weighted least square approach as shown  below in Equation \ref{eq:gxi_cal} which gives optimal value for $g_{x_i, cal}$.
\begin{equation}\label{eq:gxi_cal}
	g_{x_i, cal} =\frac{\sum_{n} C_{x_i y_k, cal} \cdot M_{k}^*}{\sum_{n} \left| M_{k}^*\right|^2} 
\end{equation}
where,
\begin{equation*}
	M_{k} = g_{y_k, cal}^*\;\;  \left[\left(d_{x_i, cal}-d_{y_k, cal}^*\right) I_{\nu,cal} \right]
\end{equation*}
\noindent The $g_{x, cal}$ obtained has to be updated for the Sun sky position to get $g_{x, sun}$ as mentioned in Section \ref{sec:calibsi}. Using $C_{x_i y_k, sun}$ obtained from observations of the Sun, $g_{x_i, sun}$ and $g_{y_k, sun}$ from the above estimations, $(d_{x_i, sun}-d_{y_k, sun}^*)$ for Sun position from the antenna simulations, and the corrected Stokes I visibilities of the Sun ($I_{\nu,sun}$), the corrected Stokes V visibilities are calculated using Equation \ref{eq:Vv_cor} shown below:
\begin{equation}\label{eq:Vv_cor}
	V_{\nu, sun (corrected)} = -j\,\frac{2\, C_{x_i y_k, sun}}{g_{x_i, sun}\;\; g_{y_k, sun}^*} +\,j\left[\left(d_{x_i, sun}-d_{y_k, sun}^*\right) I_{\nu,sun\,corrected} \right]
\end{equation}
\noindent The corrected Stokes-V visibilities are written into the UVFITS file and loaded into AIPS. Self-calibration (only for the phase) is performed and applied to get final Stokes-V visibilities of Sun and the image is obtained using the `IMAGR' task in AIPS. 

\section{Preliminary Observations and Results}\label{sec:obs_results}
\begin{figure}
	\centering
	\begin{minipage}{0.48\textwidth}
		\centering
		\includegraphics[width=\linewidth]{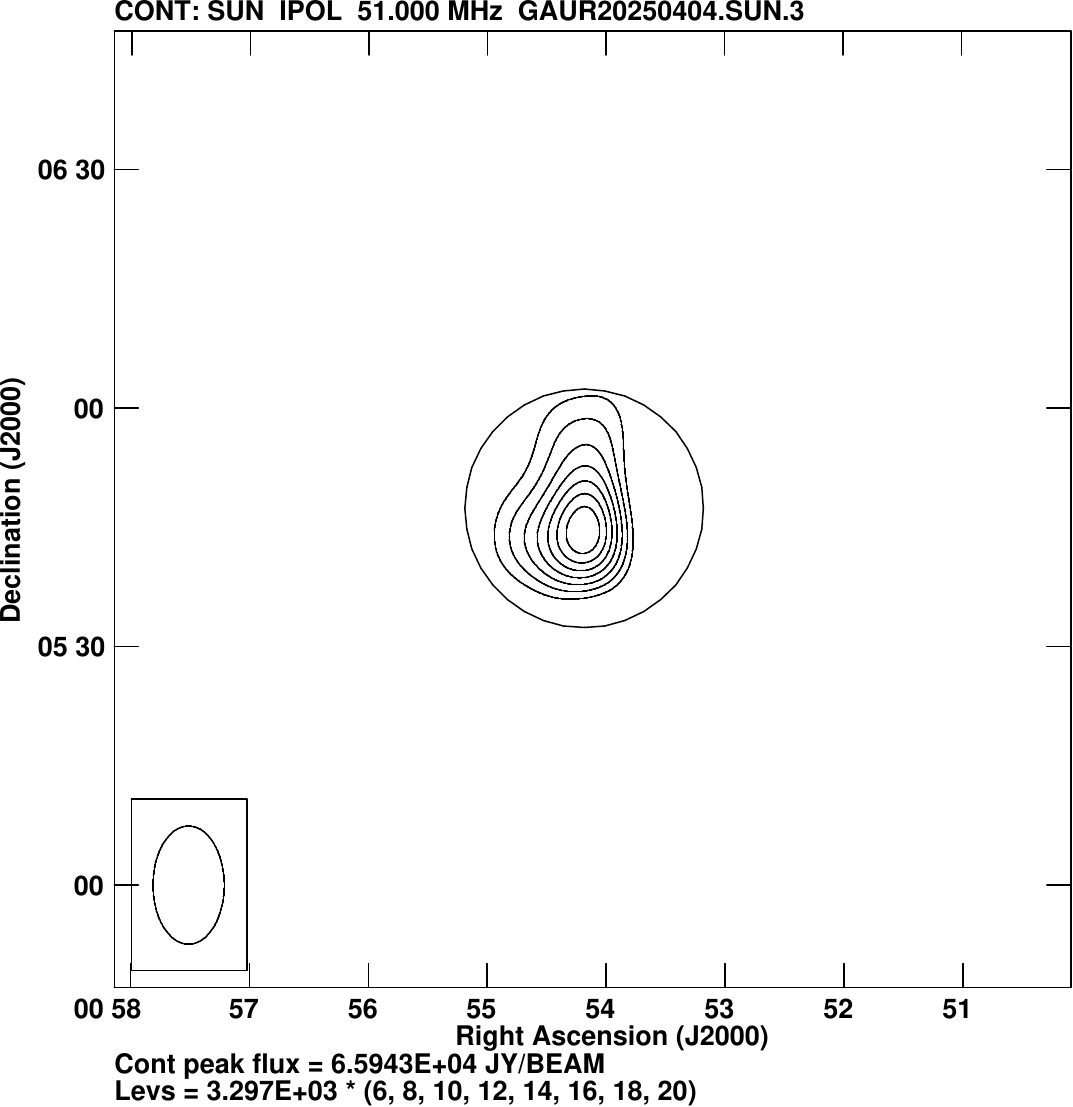}
		\\ {\footnotesize (a) Stokes I}
	\end{minipage}
	\hfill 
	\begin{minipage}{0.48\textwidth}
		\centering
		\includegraphics[width=\linewidth]{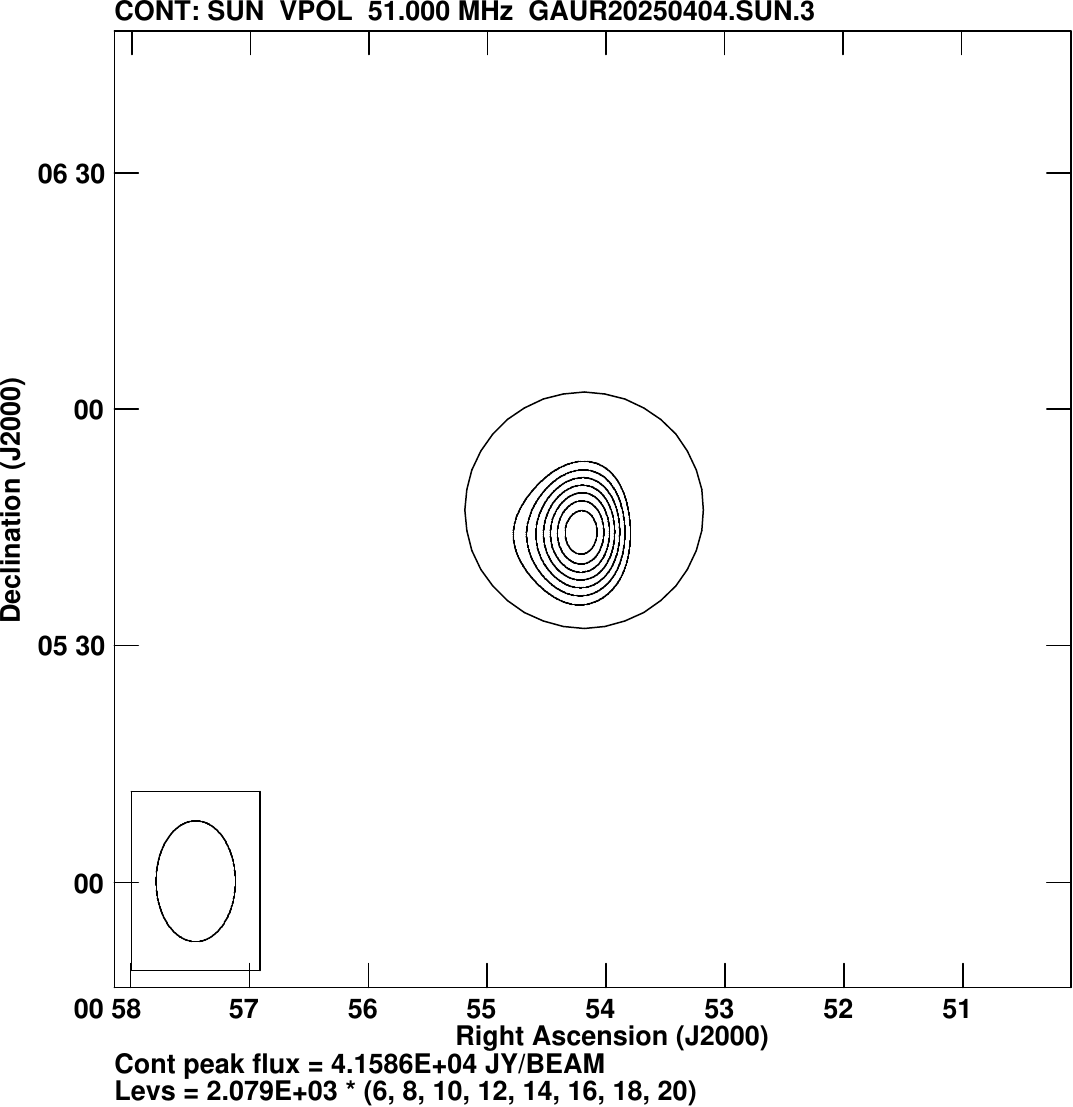}
		\\ {\footnotesize (b) Stokes V}
	\end{minipage}
	\caption{Stokes I and Stokes V images of a non-thermal noise storm continuum source observed with the augmented GRAPH at 51\,MHz on 04 April 2025. The measured DCP is 
	$\approx$\,52\%.}
	\label{fig:active_image}
\end{figure}

Trail observations of augmented GRAPH array started in the first week of April, 2025. An inspection of the solar observations with the Long Wavelength Array (LWA) in the Owens Vally Radio Observatory (OVRO) at 52\,MHz\footnote{\url{https://www.ovsa.njit.edu/lwa/}} indicates the presence of an unresolved radio source in the southern hemisphere during the epoch 1\,-\,6 April 2025. Since type I storm continuum can last several days \citep{ramesh_2011low}, it is possible that the unresolved radio source observed is due to a type I noise storm. On 01 April 2025, the GRIP also observed an intense radio emission in the 35\,-\,85 MHz band at 06:54\,UT. It appears to be non-thermal in nature because its flux density shows a decrease with increasing frequency. 
The GRIP observations during 02 April to 06 April 2025 showed that the DCP at 51\,MHz increased from $\approx$\,20\% to a maximum of ${\approx}$\,54\% on 04 April 2025, and then decreased. The DCP trend implies the sustained radio emission is most likely due to a type I noise storm continuum, as the DCP of the latter generally increases from the East limb, reaches a maximum close to the disk center, and decreases toward the west limb (see, e.g \citealp{ramesh_2011low} and the citations therein).
We used the 04 April 2025 GRIP observations as reference and assessed the performance of the augmented GRAPH. The Stokes I and V images obtained during the Sun's local meridian transit (06:54\,UT) on 04 April 2025 at 51\,MHz are shown in Figure \ref{fig:active_image}.
The corresponding peak fluxes are $6.59{\times}10^{4}$ and $4.16{\times}10^{4}$ Jy/beam, respectively. The calculated peak DCP is $\approx$\,52\%. The 52\,MHz OVRO-LWA Stokes I images showed that the peak brightness temperature of the type I continuum source is  between 2.5${\times}10^7$\,K and 8.5${\times}10^7$\,K. The corresponding GRAPH Stokes I peak brightness temperature is 6.9${\times}10^7$\,K. A comparison of the different observations mentioned above clearly indicates that the augmneted GRAPH accurately detected the total and circularly polarized radio flux densities emitted by the noise storm source in the solar corona on 04 April 2025, and the calibration method in GRAPH used is in order.

\begin{figure}
	\centering
	\begin{minipage}{0.48\textwidth}
		\centering
		\includegraphics[width=\linewidth]{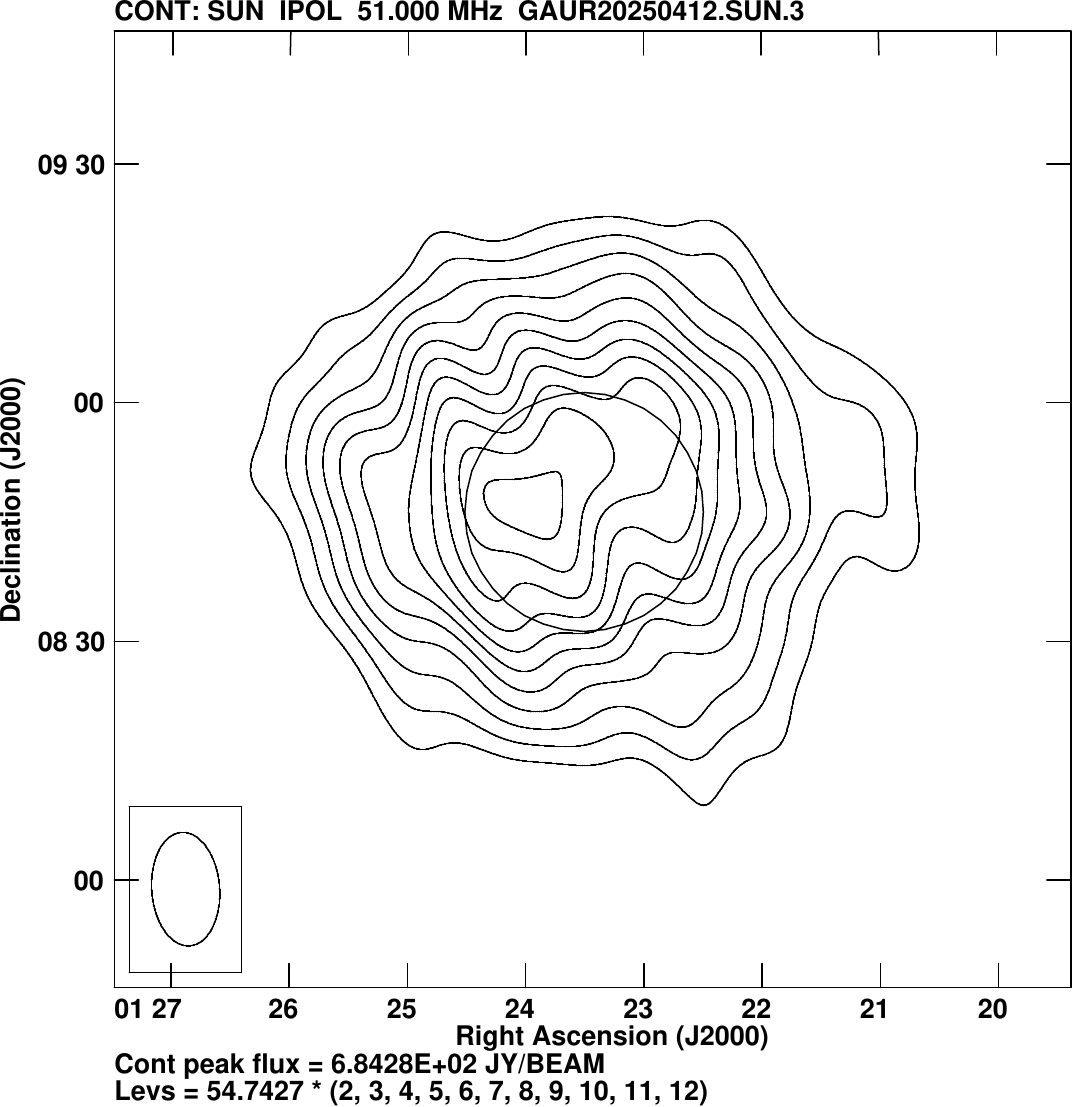}
		\\ {\footnotesize (a) Stokes I}
	\end{minipage}
	\hfill 
	\begin{minipage}{0.48\textwidth}
		\centering
		\includegraphics[width=\linewidth]{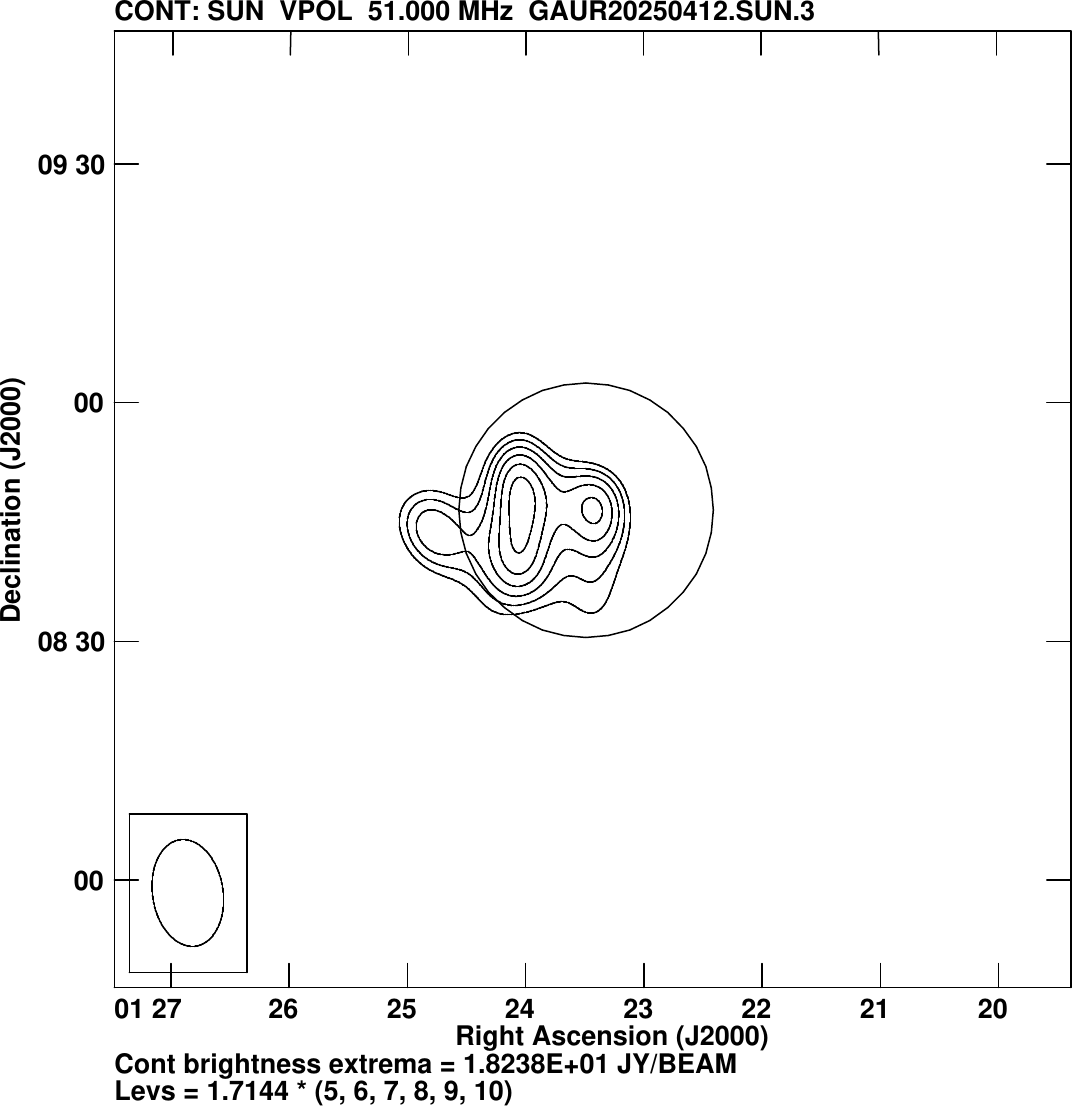}
		\\ {\footnotesize (b) Stokes V}
	\end{minipage}
	\caption{Stokes I and Stokes V images of thermal emission from the `quiet' solar corona observed with the augmented GRAPH at 51\,MHz on 12 April 2025. The observed peak DCP is $\approx$\,2.7\%.}
	\label{fig:quiet_image}
\end{figure}

Since the primary objective for the augmentation of GRAPH is to observe circularly polarized thermal emission from the `quiet' or `undisturbed' solar corona, we present GRAPH observations on 12 April 2025 at 51\,MHz (with 1\,MHz IF bandwidth and an 8-second integration time). The Sun was `quiet' during our observation time (Figure \ref{fig:quiet_image}).
The peak brightness temperature of Stokes I is $\approx$\,7.3${\times}10^5$\,K, and that of Stokes V is ${\approx}$0.2${\times}10^5$\,K. The peak temperature contours for Stokes I and Stokes V are nearly co-spatial, indicating they originate from the same localized source. The calculated peak DCP is $\approx$\,2.7\%. The OVRO-LWA\footnotemark[3] observed this thermal emission from the Sun, and the peak brightness temperature is in the range $\approx$\,6.5${\times}10^5$\,K to 7.3${\times}10^5$\,K. The Stokes I measurements with the GRAPH are consistent with those of the OVRO-LWA, similar to the case of the noise storm observations mentioned in the previous paragraph. The lowest circularly polarized flux detected is $\approx$\,7\,Jy/beam, as against the theoretical limit of about 2\,Jy for the GRAPH (Section 3.1). 

\section{Summary and Future Plans}\label{sec:summary}

The work reports the augmentation of the GRAPH array to observe circularly polarized radio emission from the solar corona, as compared to its earlier capability to observe only the total intensity. The existing GRAPH array was augmented by fabricating and installing 128 new log-periodic dipole antennas (LPDAs) in parallel with the existing LPDAs in the north-south arm of the GRAPH. The orientation of the new LPDAs are orthogonal to that of the old array, since cross-correlation of signals received by mutually orthogonal antennas help to measure circular polarization. Due to Faraday rotation, at low frequencies ($\lesssim$100\,MHz) and over a 1\,MHz operational bandwidth, any linear polarization (Stokes Q and U) originating from the solar corona can be entirely depolarized or obliterated. Consequently, Stokes Q and U parameters are assumed to be negligible, which allows to isolate total intensity (Stokes I) and circular polarization (Stokes V) signatures. Data calibration is facilitated by simulations as well as from observations of known, unpolarized strong calibrator sources. Initial observations confirm successful detection of Stokes I \& V with the augmented  GRAPH. We intend to further improve the sensitivity of GRAPH by employing direct digitization method used in modern low frequency radio telescopes elsewhere. Furthermore, we plan to use digital beam-forming technique to coherently add signals from all the antennas in each group.



%

\begin{acks}
We thank the observers and support staff in the Gauribidanur observatory for  successful completion of the GRAPH augmentation work, and regular observations. The OVSRO-LWA images mentioned in this work are due to the kind courtesy of the Owens Valley Radio Observatory Long Wavelength Array (OVRO-LWA) team, with solar observations operated by the Center for Solar-Terrestrial Research (CSTR) at the New Jersey Institute of Technology (NJIT). 
\end{acks}





%


%


%


%


%


%



%





%


%

\bibliographystyle{spr-mp-sola}

\bibliography{bibilography_paper2.bib}

%





%


%


\end{document}